\documentclass[preprint,12pt]{aastex}  %--original for ApJ

\def\cm{\ifmmode {\rm cm}^{-1} \else cm$^{-1}$ \fi}
\def\s{\ifmmode {\rm s}^{-1} \else s$^{-1}$ \fi}
\def\cc{\ifmmode {\rm cm}^{-3} \else cm$^{-3}$ \fi}
\def\cs{\ifmmode {\rm cm}^{-2} \else cm$^{-2}$ \fi}
\def\g{\ifmmode \gamma \else $\gamma$\fi}
\def\G{\ifmmode \Gamma \else $\Gamma$\fi}
\def\Gs{\ifmmode \Gamma~ \else $\Gamma~$\fi}

\def\gc{\ifmmode \gamma_{\rm c} \else $\gamma_{\rm c}$ \fi}
\def\sw{Schwarzschild~}
\def\gsim{\mathrel{\raise.5ex\hbox{$>$}\mkern-14mu
             \lower0.6ex\hbox{$\sim$}}}

\def\lsim{\mathrel{\raise.3ex\hbox{$<$}\mkern-14mu
             \lower0.6ex\hbox{$\sim$}}}

\def\simless{\mathbin{\lower 3pt\hbox
     {$\rlap{\raise 5pt\hbox{$\char'074$}}\mathchar"7218$}}}   %< or of order
\def\simmore{\mathbin{\lower 3pt\hbox
     {$\rlap{\raise 5pt\hbox{$\char'076$}}\mathchar"7218$}}}   %> or of order
\def\Msun{M_\odot}                                % solar masses
\def\4u{4U 1728--34}

\received{} \accepted{}
\slugcomment{Resubmitted to ApJ, June 27, 2007}

\lefthead{et al.} \righthead{\4u}

\shorttitle{Shock-Driven Outflows from Hot Accretion Flows}
\shortauthors{Fukumura \& Kazanas}

\begin{document}

\title{Mass Outflows from Dissipative Shocks in Hot Accretion Flows }

%\date{\today}

%
% author list tbd - include nikolai IF we use the Gamma-QPO correlation
%
\author{Keigo Fukumura \& Demosthenes Kazanas}
\affil{Astrophysics Science Division, NASA Goddard Space Flight
Center, Code 663, Greenbelt, MD 20771}
%\email{fukumura@milkyway.gsfc.nasa.gov}
\email{fukumura@milkyway.gsfc.nasa.gov, Demos.Kazanas-1@nasa.gov}
%
%\and
%
%\author{P. Reig and I. E. Papadakis}
%\affil{IESL, Foundation for Research and Technology,
%711 10 Heraklion, Crete, Greece}
%\affil{Physics Department, University of
%Crete, PO Box 2208, 710 03 Heraklion, Crete, Greece}
%\email{pau@physics.uoc.gr,jhep@physics.uoc.gr}
%
%
%

\begin{abstract}

\baselineskip=15pt

We consider stationary, axisymmetric hydrodynamic accretion flows in
Kerr geometry. As a plausible means of efficiently separating a
small population of nonthermal particles from the bulk accretion
flows, we investigate the formation of standing dissipative shocks,
i.e. shocks at which fraction of the energy, angular momentum and
mass fluxes do not participate in the shock transition of the flow
that accretes onto the compact object but are lost into collimated
(jets) or uncollimated (winds) outflows.

The mass loss fraction (at a shock front) is found to vary over a
wide range ($0\% - 95\%$) depending on flow's angular momentum and
energy. On the other hand, the associated energy loss fraction
appears to be relatively low ($\lesssim 1\%$) for a flow onto a
non-rotating black hole case, whereas the fraction could be an order
of magnitude higher ($\lesssim 10\%$) for a flow onto a
rapidly-rotating black hole. By estimating the escape velocity of
the outflowing particles with a mass-accretion rate relevant for
typical active galactic nuclei, we find that nearly 10\% of the
accreting mass could escape to form an outflow in a disk around a
non-rotating black hole, while as much as 50\% of the matter may
contribute to outflows in a disk around a rapidly-rotating black
hole. In the context of disk-jet paradigm, our model suggests that
shock-driven outflows from accretion can occur in regions not too
far from a central engine.
%
%(within 2-40 gravitational radii). Radial density profile for upstream flow
%is found to be $\sim r^{-3/2}$ while that for downstream flow can be as
%steep as $r^{-3/2}$ to $r^{-3}$.
%

Our results imply that a shock front under some conditions could
serve as a plausible site where (nonthermal) seed particles of the
outflows (jets/winds) are efficiently decoupled from bulk accretion.

\end{abstract}


\baselineskip=15pt

\section{Introduction}

%It has been widely known that the solutions to inviscid accreting
%gas around black holes may exhibit both continuous (shock-free) and
%discontinuous (shock-included) flows simultaneously.

It has been established by now that a large body of astrophysical
objects hosting supermassive black holes (e.g.,
quasars, galactic black hole candidates or microquasars) exhibit
collimated, powerful jets/winds \citep[e.g.,][for
review]{BBR84,Livio99}. In particular, strong outflows generally
occur in the radio-loud active galactic nuclei (AGNs). It is now
widely accepted that the outflows observed from a large class of
objects as jets or winds have their origin in accretion flows, which
at the same time power the radiation emission associated with these
objects \citep[e.g.,][]{BP82,Fender04}. Many astrophysical
systems apparently manage to transfer the energy from inflowing
accretion flows to outflowing jets/winds of a small population of
nonthermal particles. For instance, \cite{Junor99} in radio
observations of the nearby active galaxy M87 found a remarkably
broad jet with strong collimation occurring already at $\sim 30
-100$ \sw radii from a central engine. Recently, \citet{Kataoka07}
discussed the disk-jet connection of the radio galaxy 3C~120
observed with {\it Suzaku}. Such a population of relativistic
outflowing particles is thought to produce a subsequent synchrotron
emission that is observed in several sources
\citep[e.g.,][]{Mirabel99}. This issue clearly points to the
importance of investigating a fundamental connection between the
accreting flows and outflows in regions not too far from the central
engines -- presumably within $\sim 100$ \sw radii.
Because the emission of radiation requires the dissipation of the
kinetic energy of the accretion flow, it is not unreasonable to
suggest that the two phenomena, i.e. the dissipation/emission of
radiation and the presence of outflows, are related to the same
generic process, which manifests itself with different guises at the
diverse sites that these phenomena are observed.

Given that an accretion flow may very well become turbulent or, at a
very minimum hot due to adiabatic compression, it is not surprising
to anticipate such a flow to emit radiation, with luminosity and
spectrum depending strongly on the specific character of the
accretion process (e.g. quasi-spherical or disk). However, the
ubiquitous presence of outflows in the same objects presents a
different problem altogether: An outflow requires that a fraction of
the accreted matter be endowed with velocity higher than the escape
velocity associated with the specific radius at which the outflow is
launched. Given that the only free energy available to the accreting
gas is that of the gravitational field, hydrodynamic dissipation of
the accretion kinetic energy can never produce an outflow since this
process simply converts the kinetic energy of a gravitationally
bound flow into thermal, with the energy per unit mass (thermal plus
kinetic) never higher than that of the local gravitational
potential.

The launch of an outflow such as those observed requires the
presence of an ``engine", i.e. a mechanism that expends mechanical
work (i.e. low entropy energy) to transfer a fraction of the
available energy to an even smaller fraction of the available mass,
thereby imparting to it specific energy greater than the local
gravitational potential; it is then expected that in its further
evolution the excess energy will be converted into directed motion
and the fraction of the mass will escape to infinity in a collimated
(jet) or uncollimated (wind) outflow.

The well known models of outflows, usually involving the action of
magnetic fields such as the magnetocentrigual jet models
\citep[e.g.,][]{BP82,Contopoulos94,Konigl94,Vlahakis00}, are
specific examples as to what may constitute such an engine. These
models assume the presence of a thin Keplerian disk ``threaded" by a
poloidal magnetic field; the Keplerian rotation of the magnetic
field line footpoints and the magnetic tension transfer energy and
angular momentum to the disk plasma which can escape to infinity.
These models are consistent in that they solve simultaneously for
the poloidal field geometry and the flow velocity (under certain
simplifying assumptions, i.e. the self-similarity of the solutions).
\citet{Pelletier92} further developed the general theory of
non-self-similar solutions of hydromagnetic disk winds. In these
models the transfer of excess energy to the escaping particles is
made at the expense of the rotational energy of the matter in the
disk and it is mediated by the magnetic field.

While this type of model gained popularity, the issue of jet/outlfow
formation took a different turn with the introduction of
advection-dominated accretion flow (ADAF)
\citep[e.g.,][]{Narayan94,Manmoto97}. These radiatively inefficient
accretion flows (RIAF) were found to have positive Bernoulli
integral of the flow and could therefore fulfill the condition
necessary for the launching of jet/wind outflows; as such they
present potentially interesting sites for the origin of such
outflows. The positivity of the Bernoulli integral has been
discussed and analysed by \citet{BB99} who pointed out that it is
due to the combination of energy transfer by the viscous torques
from the inner to the outer sections of the flow (the gas of the
flow becomes bounded at its inner edge) and the local dissipation of
the flow's azimuthal kinetic energy which is not radiated away but
stored in the fluid to increase its internal energy. The latter
authors then argued that the excess energy can be carried away to
infinity (along with some fraction of the accreting mass and angular
momentum) to produce continuous outflows from all radii to infinity
while leaving the remaining flow with negative Bernoulli constant to
naturally accrete onto the compact object: advection-dominated
inflow-outflow solution (ADIOS). In this case, while the necessary
excess energy is transferred by the viscous torques from the flow's
more highly bound inner section, the necessary separation of mass to
components with positive and negative total energy is still left
unspecified.

An altogether different model that offers a simplified picture of
such a separation was presented by \citet{Subramanian99} who
proposed that in the tenuous, collisionless plasma of an ADAF
particles (protons) could be accelerated via a second-order Fermi
acceleration by the shear motions of the underlying quasi-Keplerian
azimuthal flow. They then argued that if sufficiently large pressure
is built in the accelerated particle proton population (the
electrons generally lose energy on time scales short compared to
their transit time through the system and cannot build an energy
density that could be dynamically important) and for favorable
geometries of the disk magnetic field (large scale poloidal loops
that open up above the disk) the relativistic particle population
could naturally (through the action of the gravitational field)
segregate itself from the non-relativistic one, carrying off to
infinity only the accelerated ($E \gsim m_pc^2$) portion of the disk
plasma. In this case the engine is a combination of the particle
acceleration and the action of the gravitational field.

Finally, a model along the same lines was proposed by
\citet{Contopoulos95} who suggested that even in the case of a
completely turbulent magnetic field, a separation of the
relativistic and non-relativistic particle populations is possible
through the production of relativistic neutrons in the collisions of
the relativistic protons with the ambient plasma and the ensuing
production of relativistic neutrons. The subsequent decay of
neutrons back into protons produces then a proton fluid in regions
of space devoid of inertia whose energy-to-mass ratio (and hence its
asymptotic Lorentz factor) depends only on the ratio $R/c \tau_{\rm
n}$, where $R$ is the size of the system and $\tau_{\rm n}$ the
neutron life time and can lead to highly relativistic flows for
black hole masses $M \gsim 10^8 \Msun$.

In the present note we follow a similar simplified view to study
outflows in objects powered by accretion: We consider the presence
of (2-dimensional) shocks as a means of dissipation of the accretion
kinetic energy in a fashion similar to that considered by
\citet{Cha90} and collaborators. That is we consider the transonic
accretion of matter with the proper angular momentum to produce a
standing shock at a radius close to the horizon, which, subsequently
accretes onto the black hole after passing through a downstream
sonic point. Previous steady-state analysis found that it is
possible to judiciously choose the specific angular momentum of the
flow, so that the outer transonic one could be connected through a
shock transition to an inner transonic one which, passing through an
inner sonic point, accretes onto the black hole. The location of the
shock in this situation is determined by finding a radial position
at which the density and velocity of the two flow sections were
those demanded by the dissipative Rankine-Hugoniot conditions across
a shock. So far, the formation of standing shocks in hydrodynamic
accretion has been extensively studied by a number of authors for
both inviscid flows
\citep[e.g.,][]{Cha90,Sponholz94,Cha96,Lu98,FT04} and viscous flows
\citep[e.g.,][]{Cha90,Lu99,Cha04}, considering either dissipative or
non-dissipative shock jump conditions. Recently, standing shocks in
the presence of poloidal magnetic fields [i.e., magnetohydrodynamic
(MHD) shocks] around a black hole was also studied for various
parameter dependence (see Das \& Chakrabarti 2007 for
pseudo-Newtonian geometry; Takahashi et al. 2002, Fukumura et al.
2007 for Kerr geometry). In the context of the particle acceleration
via the first-order Fermi mechanism across a shock front, the
production of shock-accelerated relativistic protons were discussed
in spherical accretion \citep[][]{Kazanas83,Kazanas86}, while other
authors have explored the relativistic outflows in ADAF with shocks
\citep[][]{Becker04,Becker05}. Similar attempts have been made to
make physical connections between the shocked-accretion and
outflows. For instance, mass outflow rate were estimated from
adiabatic shocked-flow region in Newtonian gravity
\citep[e.g.,][]{Cha99,Das00}. \citet{Das99} took a similar approach
to study the shock-generated outflows with little energy dissipation
in pseudo-Newtonian geometry. Independently, from general
relativistic MHD simulations, the formation of jets
(magnetically-driven and gas-pressure driven jets) is found in the
high pressure regions due to the shock/adiabatic compression
\citep[e.g.,][]{Koide99,Nishikawa05}. They concluded that the jets
are mainly produced by the gas-pressure gradient which is greatly
enhanced by the shock front at around $r \sim 6$ gravitational radii
(note that this feature was not seen in the Newtonian calculations).
These studies also suggest an essential connection between the
shocked accretion flows and the jets; i.e., the shock front may
serve as a base of the outflows.

The novelty of our approach lies in considering the possibility of
shock formation (i.e. obeying the general relativistic, dissipative
Rankine-Hugoniot conditions at a shock front) in which part of the
mass, angular momentum and energy fluxes escape in the z-direction
and do not participate in the shock transition. We then examine the
energy per unit mass of the escaping matter which we compare to the
escape velocity at the shock radius; if it is greater than the
latter we conclude that this scenario can produce an outflow with
its outflow rate $\dot m$ and luminosity that are calculable and can
be compared to those of the entire accretion to obtain a measure of
the efficiency of our ``engine" in producing outflows. Some
population of energetic nonthermal particles, produced via a shock
acceleration, may then be well separated from the equatorial
accretion flows (which consists primarily of thermal particles).

More specifically, since we are interested in the formation of
outflows through shocks in the inner disk region relatively close to
a presumably rotating black hole at a center (say, $r \lesssim 30$
gravitational radii), it is important to include the strong gravity
and  frame-dragging effects described by general relativity.
%Importantly, due to the fact that many black holes at the center of
%the astrophysical systems have been considered to be rotating and
%that shocks can occur in the innermost accretion disk (say, within
%ten gravitational radii), it is also important to accurately
%describe the relativistic dynamics of the accreting gas.
To the best of our knowledge, no relevant work in the literature
incorporates such mass and energy loss in the shock jump conditions.
%
%
%However, no explicit work has been done in the literature, in this
%perspective, on incorporating mass and energy loss at dissipative
%shocks in the jump conditions.
%
In the framework of our model presented here, mass loss is coupled
to energy loss via the shock jump conditions, and therefore it must
be considered simultaneously. It is this point that motivates us to
explore, for the first time, the formation of outflowing particles
as a consequence of the formation of dissipative standing shocks in
accretion in a fully relativistic treatment. The formalism of our
current model is partly based on the previous works
\citep[][]{Yang95,Lu98,FT04}. Our main objective in this study is
therefore to explore in details a possibility that the formation of
outflows (jets/winds) can occur at a dissipative shock front in
transonic accreting flows: i.e., a connection between
shocked-accreting gas and the outflowing particles.

The structure of this paper is as follows. In \S 2 we review and
explain our simplified model that simultaneously considers both
shocks and outflows with appropriate jump conditions. Main parameter
dependence of the shock-outflow solutions is explored in \S 3, where
we show the nature of the shock-outflow solutions and the
corresponding global accretion solutions. Our primary goal in these
analysis is to examine the coupling between the shock-outflow
solutions. In \S 4 we discuss our results and make some observation
implications. Brief summary and concluding remarks are given there
as well.

\section{Model Assumptions \& Basic Equations}

In black hole accretion, accreting gas must be transonic. After
passing through a first sonic radius, the gas is slowed down, and a
shock may develop. For causality, the shocked gas must become
supersonic again before crossing the event horizon. Below, we will
explain the details of our simplified model.

\subsection{Accreting Flows around a Black Hole}

We consider a steady-state, axisymmetric accreting flows in Kerr
geometry. The spacetime metric is expressed by the Boyer-Lindquist
coordinates as
\begin{eqnarray}
ds^2 &=& -\left(1-\frac{2 m r}{\Sigma}\right) dt^2 - \frac{4 m r a
\sin^2 \theta}{\Sigma} dt d\phi \nonumber \\ & & + \frac{A \sin^2
\theta}{\Sigma} d\phi^2 + \frac{\Sigma}{\Delta} dr^2 + \Sigma
d\theta^2 \ , \label{eq:BL}
\end{eqnarray}
where $\Delta \equiv r^2-2mr+a^2$, $\Sigma \equiv r^2+a^2 \cos^2
\theta$, $A \equiv (r^2+a^2)^2-a^2 \Delta \sin^2 \theta$. $m$ and
$a$ are mass and specific angular momentum (or Kerr parameter) of a
black hole. Following the standard geometrized units, we have taken
$G=c=1$ in equation~(\ref{eq:BL}). Thus, the length (distance) $r$
and $a$ are measured in units of $m$. Since we are interested in the
equatorial flows, we set $\theta=\pi/2$ throughout this paper. The
black hole horizon is then expressed by $r_{\rm{H}} \equiv
m+\sqrt{m^2-a^2}$. The self-gravity of the flow is ignored, and we
do not include the effects of magnetic fields for simplicity.

We follow the earlier works on relativistic shock formation as
follows; accretion time is assumed to be shorter than that of energy
diffusion, thus treating the flows as adiabatic except at a shock
front where a fraction of fluid energy, angular momentum, and mass
are dissipated. To prescribe thermodynamic quantities we adopt a
polytropic form as
\begin{eqnarray}
P = K \rho_0^{1+1/N} \ , \label{eq:polytropic}
\end{eqnarray}
where $P$ and $\rho_0$ are the thermal pressure and rest-mass
density of the flow, which are locally measured in the fluid frame.
Here, $N$ denotes the polytropic index. The entropy of the fluid is
characterized by $K$ which is related to the entropy $S$ by $S
\equiv c_v \log K$ where $c_v$ denotes a specific volume heat of the
flow. Because of stationarity ($\partial t=0$) and axial symmetry
($\partial \phi=0$), there exist two conserved quantities along a
fluid stream line; namely, specific energy $E$ and axial angular
momentum $L$ of the fluid, defined by
\begin{eqnarray}
E &\equiv& -\mu u_t \ , \label{eq:energy}
\\
L &\equiv& \mu u_\phi \ , \label{eq:L}
\end{eqnarray}
where $\mu = (P+\rho)/\rho_0$ is the relativistic enthalpy of the
fluid, and $\rho = \rho_0 + N P$ is the net baryon mass-energy
density (including internal energy $N P$). The number density of the
constituent baryon $n$ is given by $\rho_0 \equiv n m_p$ where $m_p$
is the baryon mass in the flow. We assume that energy $E$ and
angular momentum $L$ are both conserved along the flow except at a
shock location ($r=r_{\rm{sh}}$) where they are partially dissipated
(will be explained in detail in the next section).

From the four-velocity normalization of the flow ($u_\alpha u^\alpha
=-1$), we get
\begin{eqnarray}
1+u_r u^r + (u^t)^2 V_{\rm{eff}}(r,\lambda)=0 \ , \label{eq:V}
\end{eqnarray}
where the effective potential has been introduced by
$V_{\rm{eff}}(r,\lambda) \equiv g^{tt}- 2\lambda g^{t
\phi}+\lambda^2 g^{\phi\phi}$ with $g^{\alpha \beta}$ being the
inverse metric components for $g_{\alpha \beta}$ in
equation~(\ref{eq:BL}). Here, the specific angular momentum of the
flow $\lambda$ is defined by
\begin{eqnarray}
\lambda \equiv \frac{L}{E} = -\frac{u_\phi}{u_t} \ , \label{eq:ell}
\end{eqnarray}
which is assumed to be conserved along the whole flow in our model.
%This assumption in general does not have to be imposed, and in fact
%it may not always valid in the actual astrophysical accretion.
%However, our main purpose of this work is to systematically explore
%a characteristic nature of the energy and mass loss associated with
%the shock formation. Without the above assumption,
From equation~(\ref{eq:V}) we obtain
\begin{eqnarray}
u_t(r,\lambda) = \left[\frac{1+u_r u^r}{-V_{\rm
eff}(r,\lambda)}\right]^{1/2} \ . \label{eq:ut}
\end{eqnarray}

From the definition of enthalpy and the polytropic relation given by
equation~(\ref{eq:polytropic}) we can rewrite the enthalpy as
\begin{eqnarray}
\mu = 1+(N+1)K \rho_0^{1/N} \ . \label{eq:mu}
\end{eqnarray}
Local adiabatic sound speed $c_s$ is defined as
\begin{eqnarray}
c_s^2 \equiv \left(\frac{\partial P}{\partial \rho}\right)_{\rm{ad}}
= \frac{(1+1/N) P}{\mu \rho_0} \ . \label{eq:sound}
\end{eqnarray}
%We define flow's Mach number as
%\begin{eqnarray}
%{\cal{M}} \equiv \frac{v_{\rm CRF}}{c_s} \ , \label{eq:Mach}
%\end{eqnarray}
%where $v_{\rm CRF} \equiv (u_r u^r)/(1+u_r u^r)$ is the flow speed
%measured in the corotating reference frame (CRF).
Combining the equations~(\ref{eq:polytropic}), (\ref{eq:mu}) and
(\ref{eq:sound}) we can express $\mu$ in terms of the sound speed
$c_s$ as
\begin{eqnarray}
\mu = \frac{1}{1-N c_s^2} \ . \label{eq:mu2}
\end{eqnarray}
Accordingly, the energy of the flow $E$ in
equation~(\ref{eq:energy}) can be explicitly rewritten as a function
of $c_s$
\begin{eqnarray}
E = \left(\frac{1+u_r u^r}{-V_{\rm eff}}\right)^{1/2} / \left(1-N
c_s^2\right) \ , \label{eq:energy2}
\end{eqnarray}
which is conserved along the flow except at a shock front. That is,
across a shock front, the energy $E$ and the angular momentum $L$
will both decrease in such a way that the ratio $\lambda \equiv L/E$
is continuous across the shock. Using equation~(\ref{eq:mu}) and
(\ref{eq:mu2}) the baryon rest-mass density can be rewritten as
\begin{eqnarray}
\rho_0 = \left[\frac{c_s^2}{\left(1+1/N\right)\left(1-N
c_s^2\right)}\right]^N \frac{1}{K^N} \ . \label{eq:rho}
\end{eqnarray}
We define the mass-accretion rate $\dot{M}$ as
\begin{eqnarray}
\dot{M} \equiv -4\pi r H u^r \rho_0 \ , \label{eq:mdot1}
\end{eqnarray}
where $H$ represents the vertical scale-height of the flow defined
as
\begin{eqnarray}
H \equiv \frac{c_s}{\Omega_K} \ , \label{eq:H}
\end{eqnarray}
from the conventional hydrostatic equilibrium assumption. Here,
$\Omega_K(r) \equiv m^{1/2}/(r^{3/2}+a m^{1/2})$ is the Keplerian
angular velocity. Note that for accretion we have $u^r<0$. As often
assumed, we only consider a constant mass-accretion rate in this
paper, although a variable accretion rate has been discussed in the
literature \citep[e.g.,][]{BB99}. Eliminating $\rho_0$ from
equations~(\ref{eq:rho}) and (\ref{eq:mdot1}) we rewrite $\dot{M}$
as
\begin{eqnarray} \dot{M} = -4\pi r
m^{1/2} \left(r^3/2+a m^{1/2}\right) u^r
\left[\frac{c_s^{2N+1}}{(1+1/N)(1-N c_s^2)^N}\right] \frac{1}{K^N} \
. \label{eq:mdot2}
\end{eqnarray}
Similarly to $E$ and $L$, the mass-accretion rate $\dot{M}$ is also
conserved along the flow except at a shock location. After defining
all the physical quantities necessary to solve for black hole
accretion, we will describe below the transonic properties of the
physical accretion solutions.

\subsection{Regularity Conditions}

For accretion to continue on to the event horizon, it is required
that the accreting flows become supersonic at a sonic radius. After
taking the derivatives of the equations~(\ref{eq:energy2}) and
(\ref{eq:mdot2}) with respect to $r$ (note that $E$ and $\dot{M}$
are both constants), $dc_s/dr$ can be eliminated. Finally we obtain
\begin{eqnarray}
\frac{du^r}{dr} = \frac{\cal{N}}{\cal{D}} \ , \label{eq:critical}
\end{eqnarray}
where
\begin{eqnarray}
{\cal{D}} &\equiv& 2N c_s^2 + u_r u^r \left\{N \left(3c_s^2-2
\right)-1 \right\} \ , \label{eq:D}
\\
{\cal{N}} &\equiv& \zeta \left(1+u_r u^r \right) \left(r^{3/2}+a
m^{1/2}\right) \frac{dV_{\rm eff}}{dr} \nonumber \\ & & - \left[a
\left(4Nc_s^2+\eta \right)+r^{3/2} \left\{2Nc_s^2
\left(5+3g_{rr}\right)+\eta \right\} \right] V_{\rm eff} \ ,
\label{eq:N}
\end{eqnarray}
and
\begin{eqnarray}
\eta &\equiv& 4N g_{rr} c_s^2 + \zeta (u^r)^2 \frac{d g_{rr}}{dr} \
. \label{eq:eta}
\\
\zeta &\equiv& r \left\{N \left(c_s^2-2 \right)-1 \right\} \ .
\label{eq:zeta}
\end{eqnarray}
In equation~(\ref{eq:critical}), the fact that the velocity gradient
$du^r/dr$ is finite at $r=r_c$ requires that ${\cal{D}}(r=r_c)=0$
and ${\cal{N}}(r=r_c)=0$ simultaneously (i.e., regularity
conditions), which will allow us to find the critical radius $r_c$
for a given parameter set. A physically valid accretion solution,
therefore, must pass through a critical point at $r=r_c$ before
reaching the horizon whether or not the shock formation is possible.
In the presence of a shock, a global shock-included accretion
solution must go through a critical point on both sides of the shock
location (i.e., before and after the shock forms). This fact
requires multiple critical points. It is widely known that multiple
critical points (up to three at most) can exist in general for a
certain flow parameter space. Much work has been done on examining
the topological behaviors (i.e., saddle, node, center and spiral
points) of the critical points \citep[e.g.,][]{Cha90}, thus we will
not repeat this. Because our main goal of this paper is to explore a
possibility of outflows that are coupled to the shocks in the global
accretion flows, we will only investigate the upstream flows passing
through the outer critical points $r_c^{\rm out}$ while the
downstream flows passing through the inner critical points $r_c^{\rm
in}$ (the middle one is known to be unphysical).

So far, for a specified flow parameter set, one can obtain accretion
solutions. Next, let us impose the shock conditions that connects
one solution (i.e. upstream flow) to another (i.e. downstream flow),
taking into account energy, angular momentum, and mass loss at a
shock location.

\subsection{Shock Formation with Energy and Mass Loss}

Accreting flows around a black hole are generally subject to a
number of ``invisible'' obstacles that can decelerate the flow: (1)
centrifugal barrier due to the fluid's angular momentum, (2) gas
pressure-gradient, (3) radiation pressure-gradient, and (4) magnetic
forces (i.e., pressure-gradient and/or tension force). Although our
model is purely adiabatic and hydrodynamic [thus (3) and (4) are
absent], flows are still under the influence of the deceleration
mechanisms (1) and (2).

\begin{figure}[t]% ------------------------------------- Figure~1
    \centering
    \epsscale{0.5}
    \includegraphics[angle=0, width=3.5in]{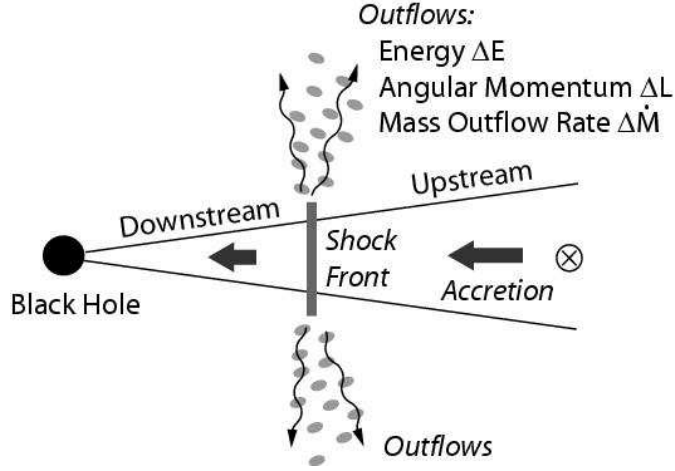}
    \caption{Schematic picture of mass outflows originating from a standing shock front
    in hydrodynamic accretion. $\bigotimes$ indicates azimuthal motion of the flow.
    At the shock, a fraction of the accreting gas is lost
    carrying energy $\Delta E$, angular momentum $\Delta L$ and mass
    outflow rate
    $\Delta \dot{M}$.
    } \label{fig:schematic}
\end{figure} %-----------------------------------------------

Following the previous works \citep[see][]{Yang95,Lu98}, let us
assume jump conditions that allow energy, angular momentum and mass
loss at a standing shock front. Figure~\ref{fig:schematic}
illustrates a schematic description of our model. From
equation~(\ref{eq:energy2}) we have
\begin{eqnarray}
E_1 &=& \left[\frac{1+g_{rr} (u_1^r)^2}{-V_{\rm eff}}\right]^{1/2} /
\left(1-N c_{s1}\right) \ , \label{eq:E1}
\\
E_2 &=& \left[\frac{1+g_{rr} (u_2^r)^2}{-V_{\rm eff}}\right]^{1/2} /
\left(1-N c_{s2}\right) \ , \label{eq:E2}
\end{eqnarray}
where the subscripts ``1'' and ``2'' denote the quantities for
upstream and downstream flows evaluated at a shock location
($r=r_{\rm sh}$), respectively. We require $E_1 > E_2$ and define
the energy dissipation and its fraction as
\begin{eqnarray}
\Delta E \equiv E_1 - E_2  ~~~~  {\rm and} ~~~~ f_E \equiv
\frac{\Delta E}{E_1} \ , \label{eq:DeltaE}
\end{eqnarray}
where $0 < f_E < 1$. The associated angular momentum carried by the
outflows is then given by $\Delta L \equiv L_1 - L_2 = \lambda
\Delta E$.

In the baryon mass conservation across the shock front, we also
consider some mass loss (equivalently the loss of mass-accretion
rate) associated with the outflows that are blown away as
winds/jets. From equation~(\ref{eq:mdot2}) we have
\begin{eqnarray}
\dot{M}_1 = -4\pi r_{\rm{sh}} \left(r_{\rm{sh}}^3/2+a m^{1/2}\right)
u_1^r \left[\frac{c_{s1}^{2N+1}}{(1+1/N)\left(1-N
c_{s1}^2\right)^N}\right] \frac{1}{K_1^N} \ , \label{eq:Mdot1}
\\
\dot{M}_2 = -4\pi r_{\rm{sh}} \left(r_{\rm{sh}}^3/2+a m^{1/2}\right)
u_2^r \left[\frac{c_{s2}^{2N+1}}{(1+1/N)(1-N c_{s2}^2)^N}\right]
\frac{1}{K_2^N} \ . \label{eq:Mdot2}
\end{eqnarray}
Clearly, the mass-accretion rate depends on the measurement of
entropy $K$ which must increase across the shock because of the heat
generated (second law of thermodynamics). However, in the presence
of both energy dissipation and mass loss at the shock front, a
fraction of total energy (including mass and thermal energies) can
be released from the flow surface, quickly reducing the rise of $K$.
This hypothesis is justifiable when cooling processes are very
efficient, although it is beyond the scope of this work to discuss
the details of these mechanisms. Hence, for simplicity, we assume
that the entropy will roughly remain unchanged at the shock front as
a result from partial heat loss of the shocked flow. Thus, we set
$K_1 = K_1 \equiv K_0$ and require $\dot{M}_1 \ge \dot{M}_2$.
Similarly to energy dissipation, let us define the mass loss and its
fraction as
\begin{eqnarray}
\Delta \dot{M} \equiv \dot{M}_1 - \dot{M}_2 ~~~~  {\rm and} ~~~~
f_{\dot{M}} \equiv \frac{\Delta \dot{M}} {\dot{M}_1} \ ,
\label{eq:DeltaMdot}
\end{eqnarray}
where $0 \le f_{\dot{M}} < 1$. That is, $f_{\dot{M}}=0$ corresponds
to no mass outflows from shocks.

The momentum flux density $T^{\alpha \beta}$ for an ideal fluid is
given by
\begin{eqnarray}
T^{\alpha \beta} = (\rho+P)u^{\alpha}u^{\beta} + P g^{\alpha \beta}
\ . \label{eq:momentum}
\end{eqnarray}
Therefore, the radial component $T^{rr}$ is
\begin{eqnarray}
T^{rr} = \rho_0 \mu u^r \left[u^r+\frac{c_s^2}{(1+1/N)u_r}\right] \
. \label{eq:Trr}
\end{eqnarray}
From the mass and momentum conservations in radial direction, we
finally obtain
\begin{eqnarray}
\mu_1 c_{s2} \left[u_1^r + \frac{c_{s1}^2}{(1+1/N)g_{rr}
u_1^r}\right] = \mu_2 c_{s1} \left(1-f_{\dot{M}}\right) \left[u_2^r
+ \frac{c_{s2}^2}{(1+1/N)g_{rr} u_2^r}\right] \ ,
\label{eq:mom-cons}
\end{eqnarray}
which completes a series of our dissipative shock conditions.

The strength of shocks is measured by the local compression ratio of
the flow, $n_2/n_1$, which is expressed as
\begin{eqnarray}
\frac{n_2}{n_1} = \frac{u^r_1}{u^r_2} \frac{c_{s1}}{c_{s2}}
(1-f_{\dot{M}}) = \left[ \left(\frac{c_{s2}}{c_{s1}}\right)^2
\left(\frac{1-N c_{s1}^2}{1-N c_{s2}^2}\right) \right]^N \ .
\label{eq:n12}
\end{eqnarray}
Hence, in the absence of mass loss ($f_{\dot{M}}=0$), the
compression ratio must be greater than unity when shocks occur.
However, in the presence of mass loss ($0 < f_{\dot{M}} <1$), the
compression ratio becomes a product of the velocity ratio and the
mass loss fraction. Therefore, there can be a case where rarefaction
or decompression with $n_2/n_1 \le 1$ might take place even if the
fluid velocity (and sound speed) abruptly decreases across a very
strong shock. To avoid further complications, we will focus our
attention on compression shock waves only.

\subsection{Dynamical Stability of Shocks}

In terms of the radial momentum balance across the shock front, some
standing shocks can be dynamically unstable. That is, the shock may
decay away (either radially inward or outward) as a result of a
small (radial) perturbation of its position. In order to examine the
stability of the obtained shock solutions, we perturb the radial
momentum flux density $T^{rr}$ (equivalent to pressure) by invoking
an infinitesimally small variation of the shock location $\delta
r_{\rm sh}$. If a shock front shifts back to its original location
to retain the momentum equilibrium there, it is dynamically stable.
The criteria is expressed as
\begin{eqnarray}
\left(\delta T^{rr}_2 - \delta T^{rr}_1\right)_{\rm sh} =
\left(\frac{d T^{rr}_2}{dr} - \frac{d T^{rr}_1}{dr}\right)_{\rm sh}
\delta r_{\rm sh} \equiv \kappa(r_{\rm sh}) \delta r_{\rm sh} \ ,
\label{eq:stability}
\end{eqnarray}
where $\kappa(r_{\rm sh})$ is a function of $r_{\rm sh}$ alone, thus
possible to be numerically evaluated. In our calculations we take
advantage of the known fact that $\kappa(r_{\rm sh})<0$ always
guarantees the stable standing shocks
\citep[see][]{Yang95,Lu98,FT04}.

\section{Numerical Results}

Following the formalism outlined above we calculate the fractions of
energy and mass losses for various model parameters. There are
essentially three primary variables ($E_1,\lambda;r_{\rm sh}$) that
determine the solutions for a given geometry ($a,\theta$). In this
paper, we restrict ourselves to the flows corotating with the black
hole ($a \lambda > 0$) at the equator ($\theta=\pi/2$).

The topology of the accreting solutions is normally classified as
``x-type" or ``$\alpha$-type" depending on whether the shock-free
solution is global or not \citep[e.g., see][for definition]{Lu98}.
Although this is also one of the important aspects of the studies of
standing shocks, we do not consider the distinction here since it is
not crucial to our current investigations.

In our calculations we set $N=3$ and choose several representative
values of the upstream flow energy: $E_1=1.003,1.004,1.005$ for
$a/m=0$ (\sw case) and $0.99$ (Kerr case) to illustrate the
frame-dragging effect.
%We
%have explored the parameter space that allows both dissipative
%shocks and mass loss, from which the values have been chosen in our
%calculations.

%It should be noted that for a given $\lambda$ there exists the
%degeneracy of shock-outflow solutions: i.e., for a single value of
%angular momentum $\lambda$ there allowed a finite range of the
%possible shock locations $r^{min}_{sh} < r_{\rm sh} < r_{\rm sh}^{\rm max}$
%corresponding to respective outflows $f_{\dot{M}}^{min} \le
%f_{\dot{M}} \le f_{\dot{M}}^{max}$ where the indices ``min" and
%``max" denote minimum and maximum values, respectively.
%
%
%In the context of our model, one can not uniquely remove this
%degeneracy, although it should be removed by additional physical
%conditions (e.g., viscosity, magnetic fields, for instance).
%

\subsection{Dependence of Energy and Mass Loss}

\begin{figure}[t]% ------------------------------------- Figure~2
\epsscale{0.3} \plotone{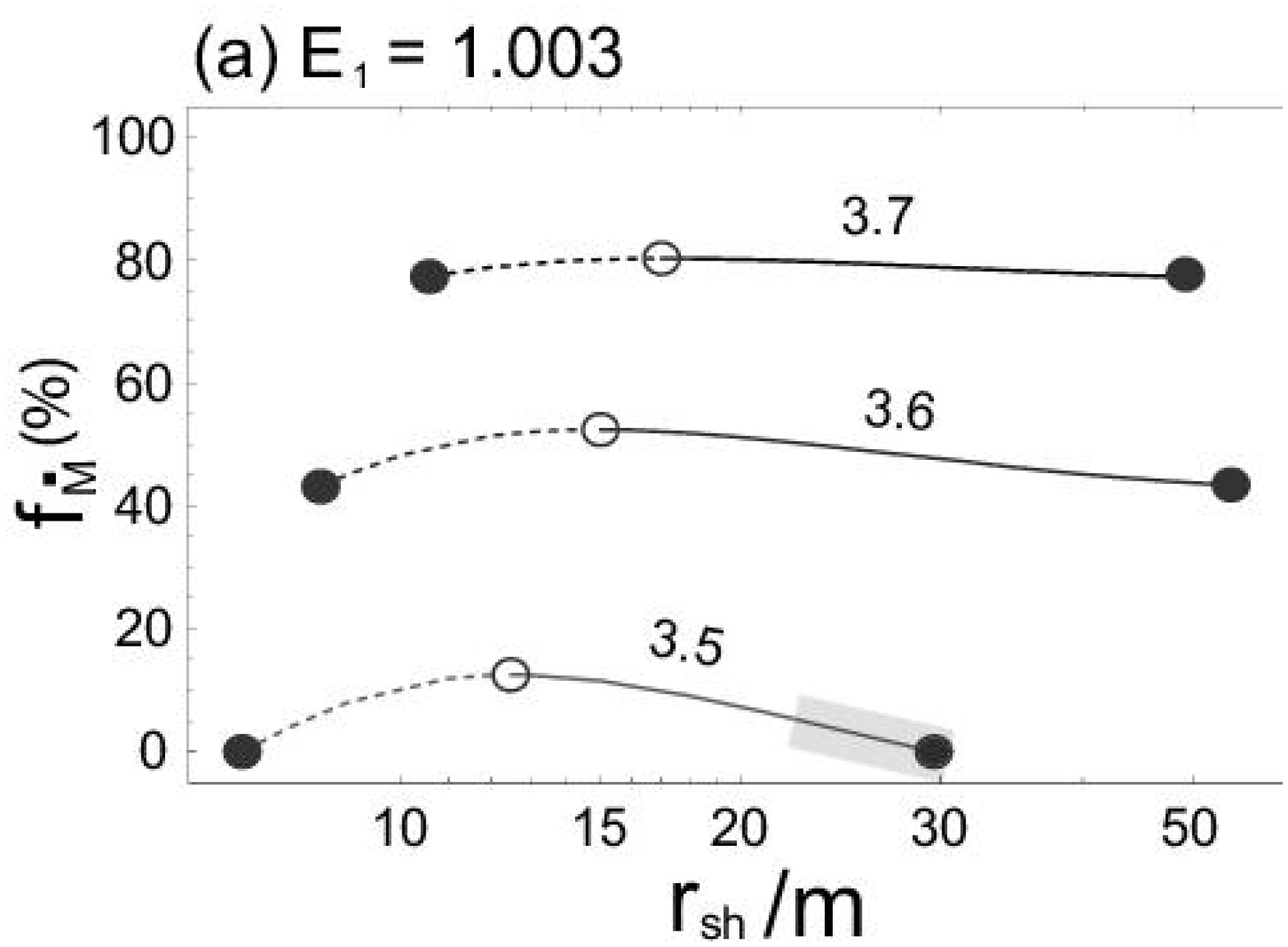}\plotone{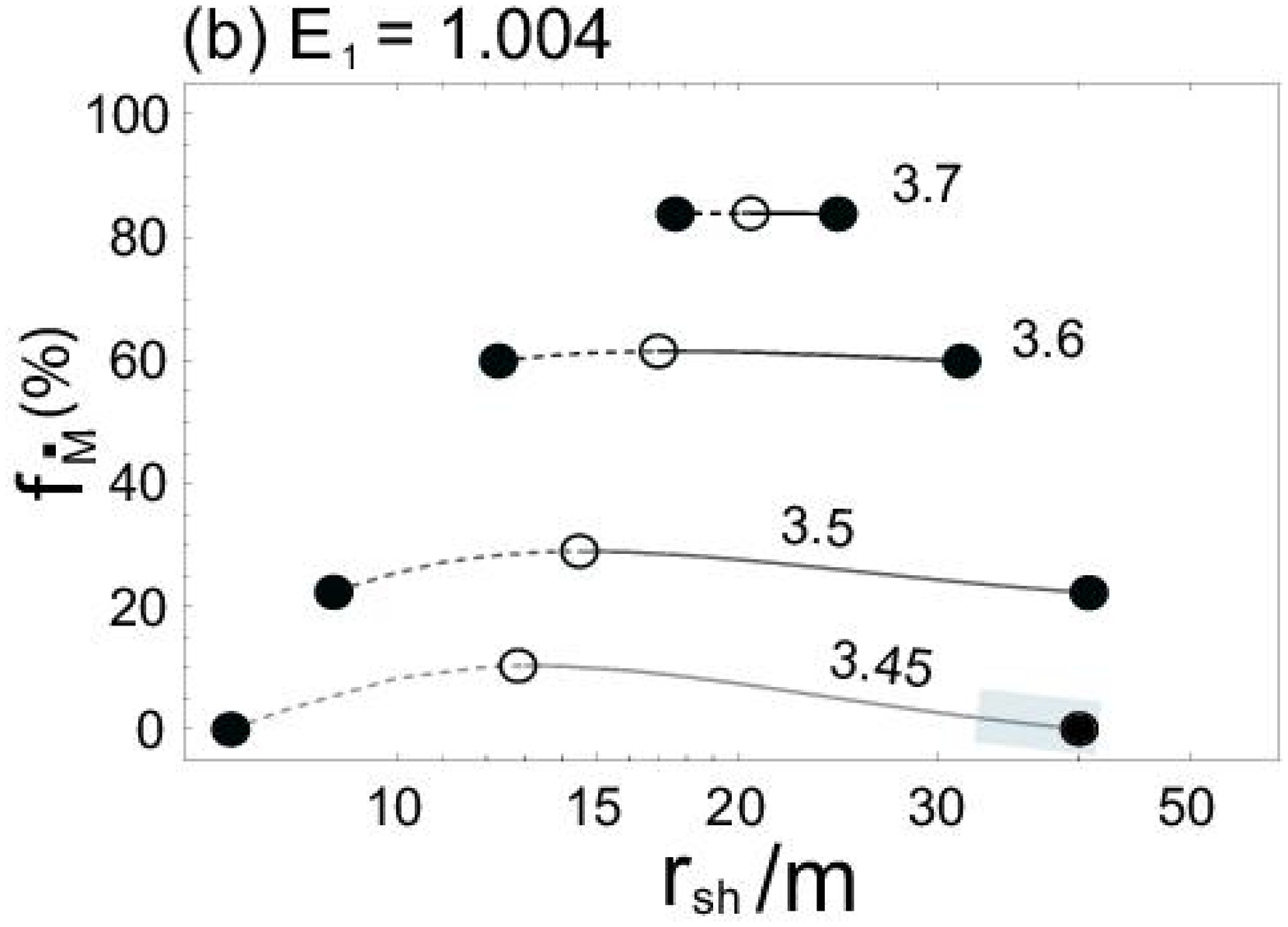}\plotone{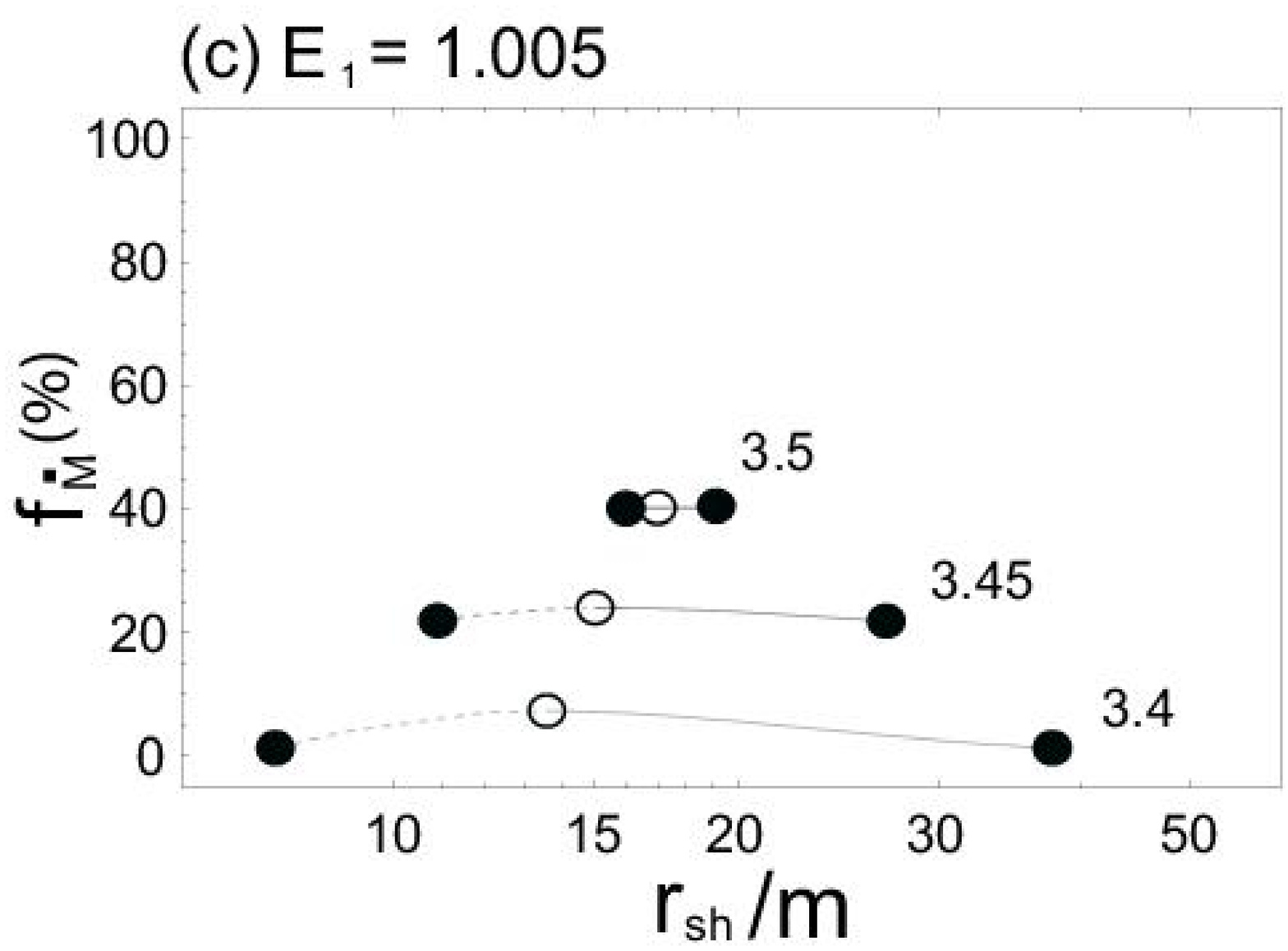}
\caption{$f_{\dot{M}}$ vs. $r_{\rm sh}$ for various flow energies
(a) $E_1=1.003$, (b) 1.004, and (c) 1.005 with $a/m=0$. Stable
(unstable) shocks are denoted by solid (dotted) curves. In each
branch, filled circles denote the maximum stable shock location also
corresponding to the weakest shock (smallest $n_2/n_1$), while open
circles denote the minimum stable shock location also corresponding
to the strongest shocks (largest $n_2/n_1$). A labeled value of
angular momentum $\lambda$ is fixed for each curve. Unbound outflow
solutions are indicated by shaded boxes. } \label{fig:rsh-a0}
\end{figure}

\begin{figure}[t]% ------------------------------------- Figure~3
\epsscale{0.3} \plotone{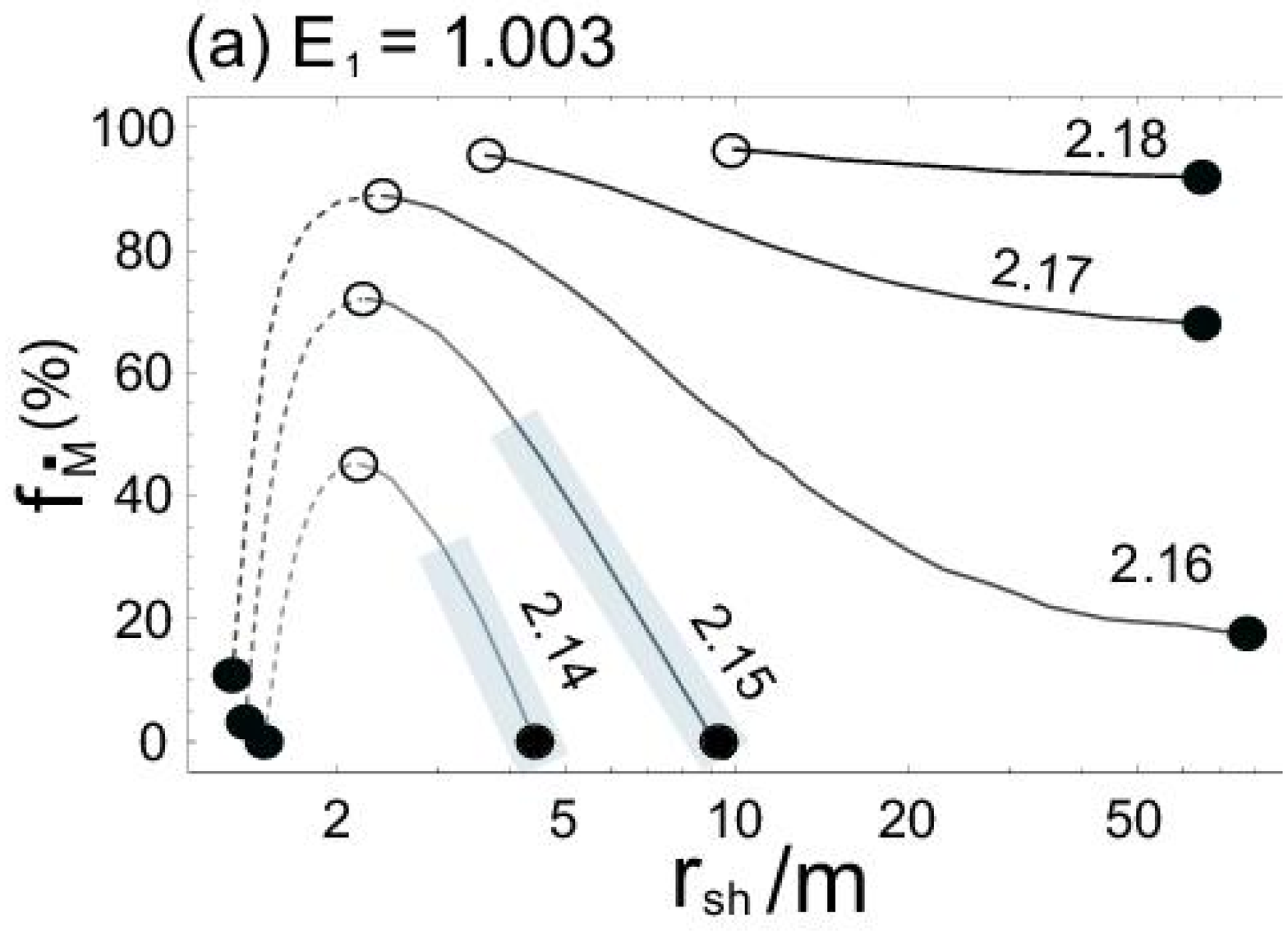}\plotone{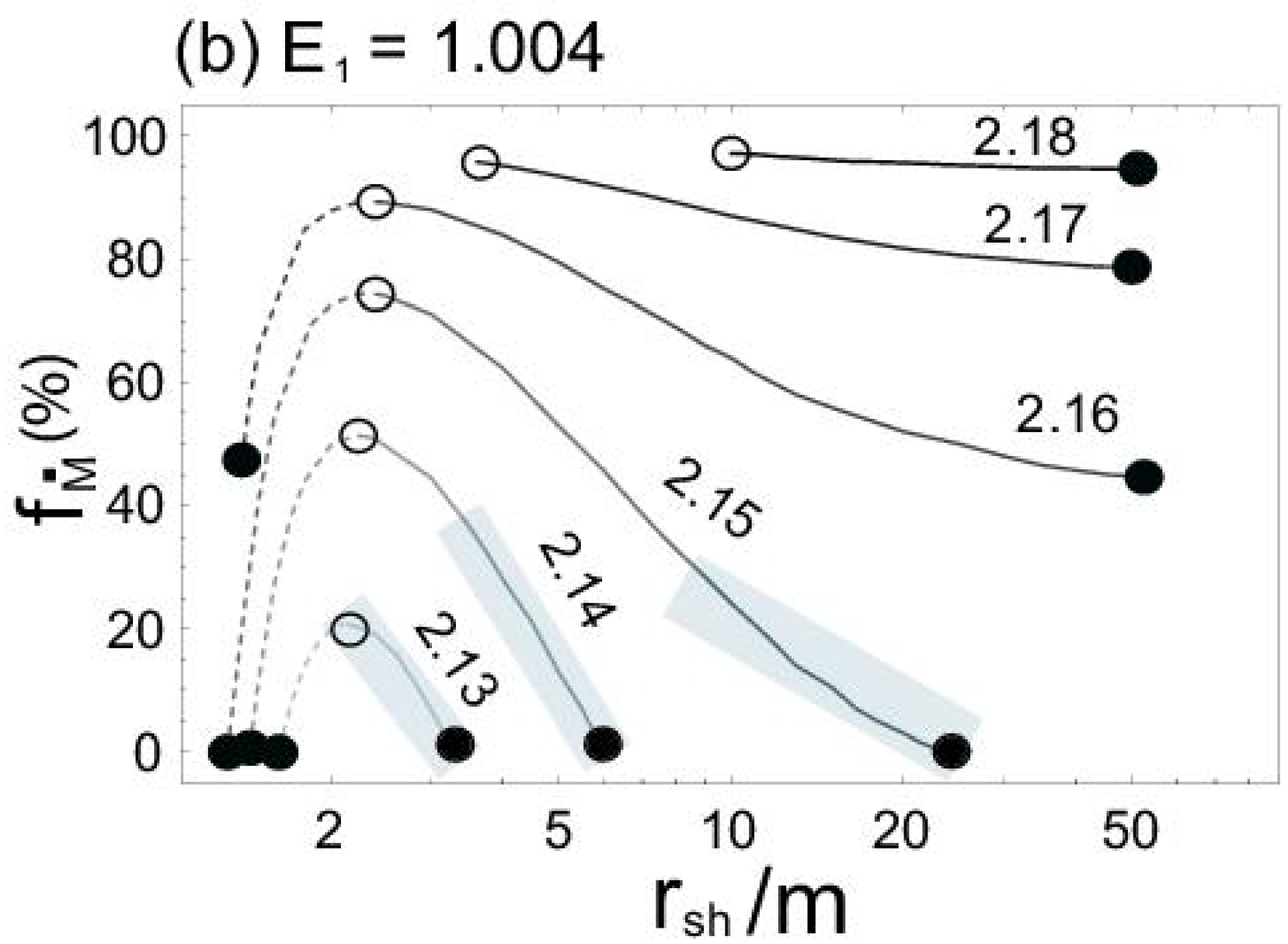}\plotone{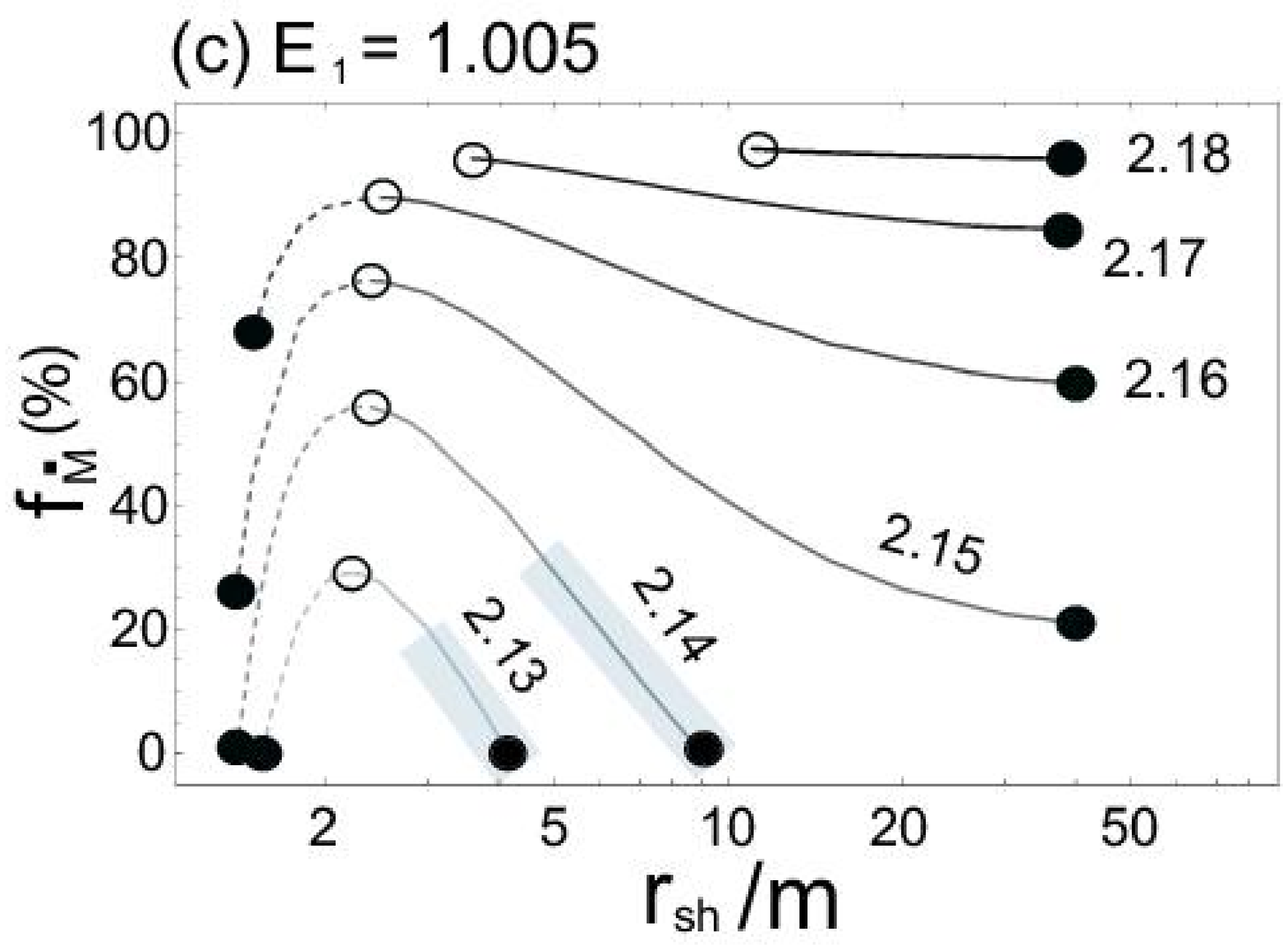}
\caption{Same as Fig.~\ref{fig:rsh-a0} but for $a/m=0.99$ case.}
\label{fig:rsh-a099}
\end{figure}

We first present in Figures~\ref{fig:rsh-a0} and \ref{fig:rsh-a099}
the mass loss fraction $f_{\dot{M}}$ as a function of shock location
$r_{\rm sh}$ for (a) $E_1=1.003$, (b) 1.004, and (c) 1.005, with
$a/m=0$ (Figure~\ref{fig:rsh-a0}) and $0.99$
(Figure~\ref{fig:rsh-a099}). Stable shocks are represented by solid
curves while (dynamically) unstable shocks are represented by dotted
curves. Stable standing shocks, according to the criterion of
equation~(\ref{eq:stability}), can form in most cases in regions
relatively close to the black hole ($r_{\rm sh}/m \lesssim 80$). For
a rotating black hole case, the shock location can be considerably
closer to the hole ($r_{\rm sh}/m \gtrsim 2-3$), results similar to
no mass loss cases \citep[e.g.,][]{Sponholz94,Lu98}, due to the fact
that the horizon shifts more inward. There appears to be a smooth
transition between the stable and unstable ones. Filled circles
denote the {\it maximum} stable shock location, also corresponding
to the {\it weakest} shock (i.e., smallest $n_2/n_1$), while open
circles denote the {\it minimum} stable shock location, also
corresponding to the {\it strongest} shocks (i.e., largest
$n_2/n_1$). In other words, the stronger dissipative shocks can
develop at smaller radii, in agreement with the previous result in
the absence of mass loss \citep[e.g.,][in which
$f_{\dot{M}}=0$]{Lu98,FT04}. Unstable shocks start to develop from
where stable shocks become strongest. At both ends of the solution
the curves are restricted by the dissipative shock conditions
[equations~(\ref{eq:E1}), (\ref{eq:E2}), (\ref{eq:Mdot1}),
(\ref{eq:Mdot2}), (\ref{eq:mom-cons})] and the transonic properties
[equation~(\ref{eq:critical})]. It is seen that smaller angular
momentum is required for shock (and accretion) to take place when
the fluid energy is larger, also consistent with previous studies of
standing shocks \citep[][]{Cha96,Lu97,Lu98,FT04}. Only stable shocks
alone are allowed when angular momentum becomes sufficiently large
(see Fig.~\ref{fig:rsh-a0}c and Fig.~\ref{fig:rsh-a099}). To
simplify our discussion, we will focus primarily on stable shocks
alone from this point. As the angular momentum of the flow $\lambda$
increases, the shock location tends to (but not always) shift
radially outward for a given energy $E_1$ as expected because the
centrifugal force correspondingly increases at a given radius.
Hence, accreting flow subject to more outward force must decelerate
at larger distance, forcing the shock location to shift outward, as
seen in our results. This trend appears to be more clear for the
$a/m=0.99$ case. It is also noted that in these figures larger
angular momentum can allow for a larger mass outflow fraction
$f_{\dot{M}}$.

In our model, mass loss fraction $f_{\dot{M}}$ can vary over a wide
range ($0\% \lesssim f_{\dot{M}} \lesssim 95\%$) depending on the
shock location and angular momentum. That is, mass outflows are not
suppressed by relativistic effects and energy dissipation \citep[for
comparion with psuredo-Newtonian case with no energy dissipation,
see, e.g.][]{Das99}. For a fixed angular momentum, on the other
hand, a higher value of $f_{\dot{M}}$ is expected from stronger
shocks occurring in the inner regions, and this seems to be the case
more in rotating black hole cases (see Fig.~\ref{fig:rsh-a099}). We
shall explain the shaded regions in these figures in the Discussion
section.

 % -------------------------------------

 % -------------------------------------

\begin{figure}[t]% ------------------------------------- Figure~4
\epsscale{0.3} \plotone{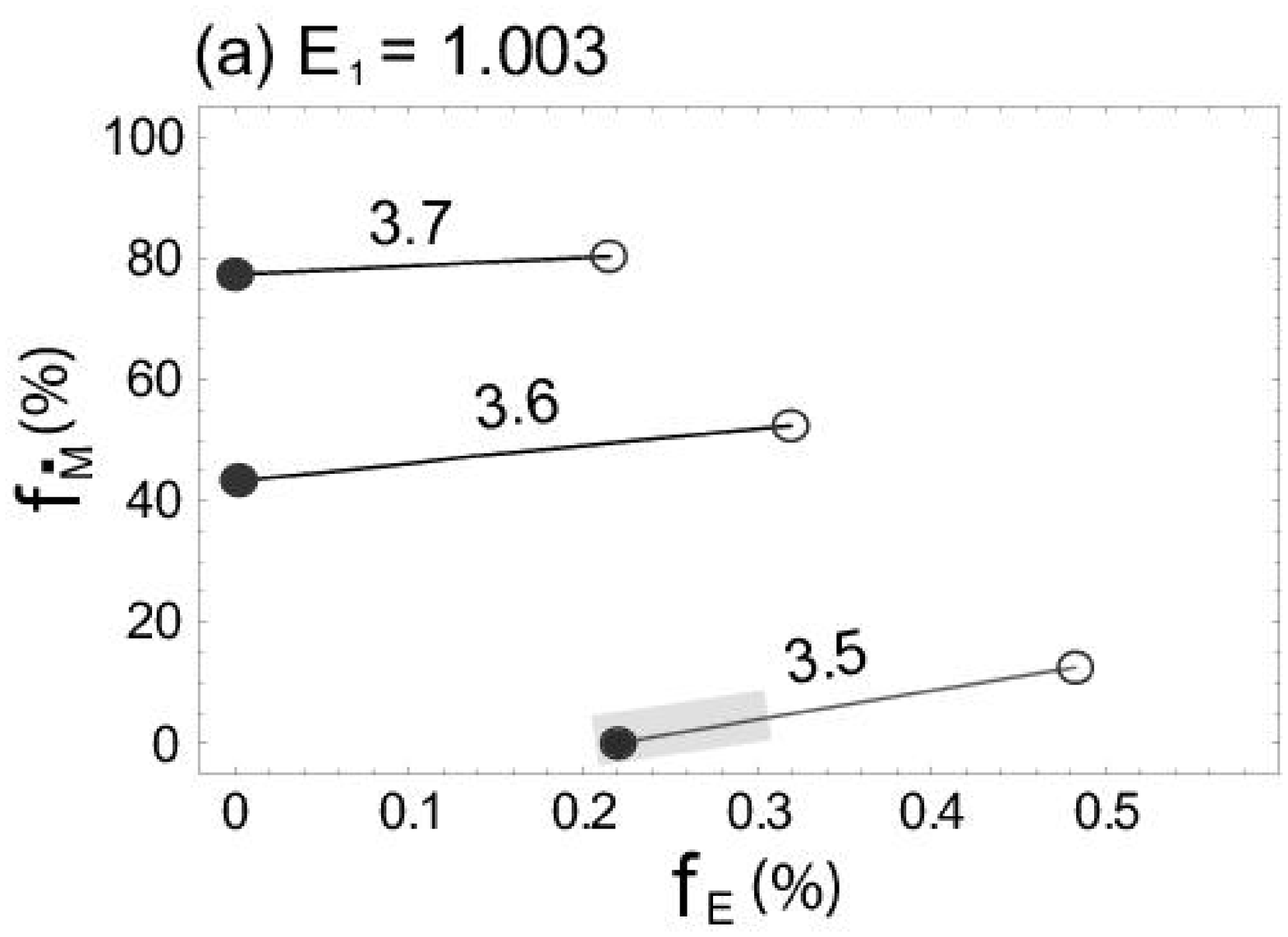}\plotone{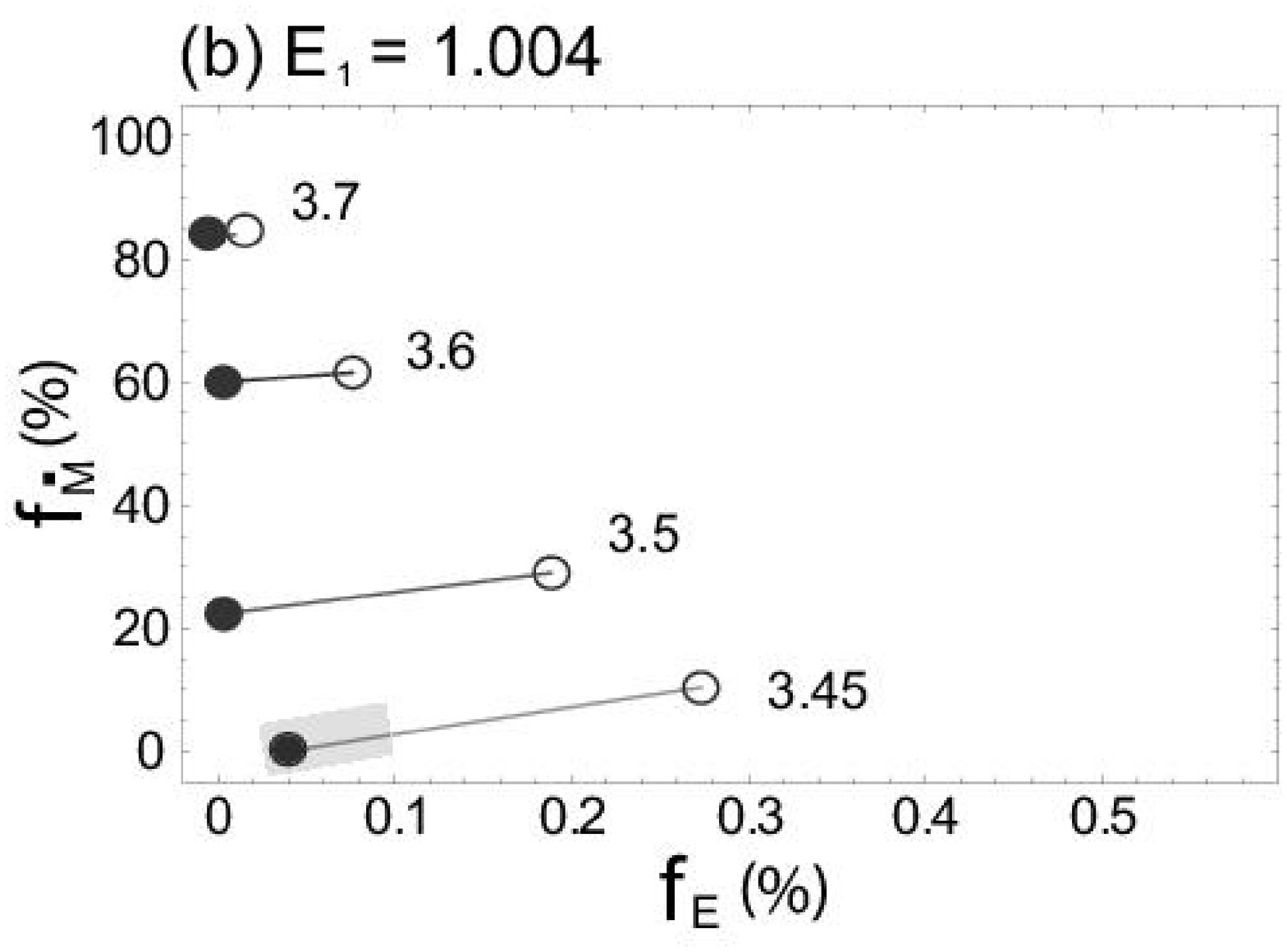}\plotone{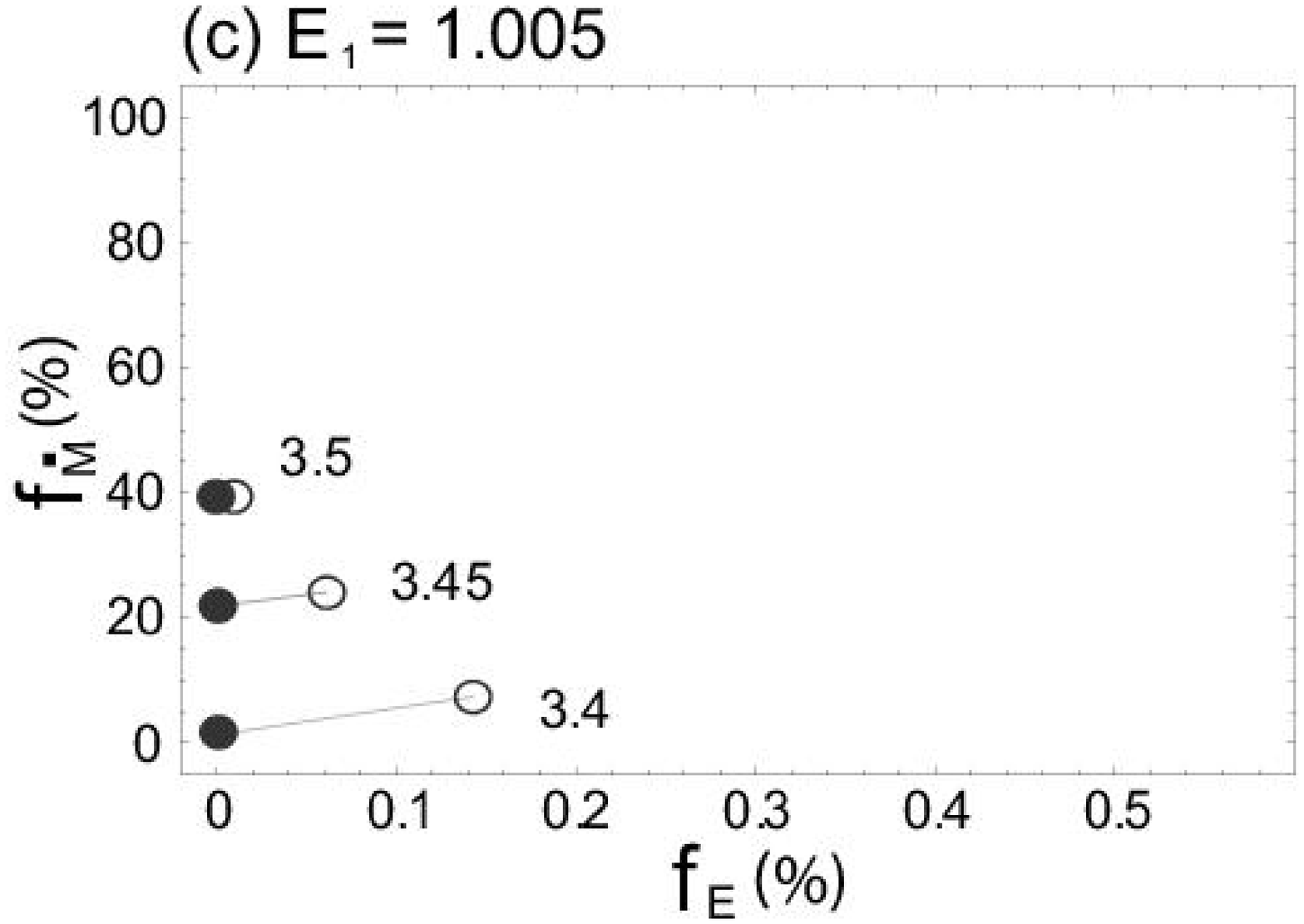}
\caption{$f_{\dot{M}}$ vs. $f_E$ corresponding to the solutions in
Fig.~\ref{fig:rsh-a0} for $a/m=0$ case. } \label{fig:E-a0}
\end{figure}

\begin{figure}[t]% ------------------------------------- Figure~5
\epsscale{0.3} \plotone{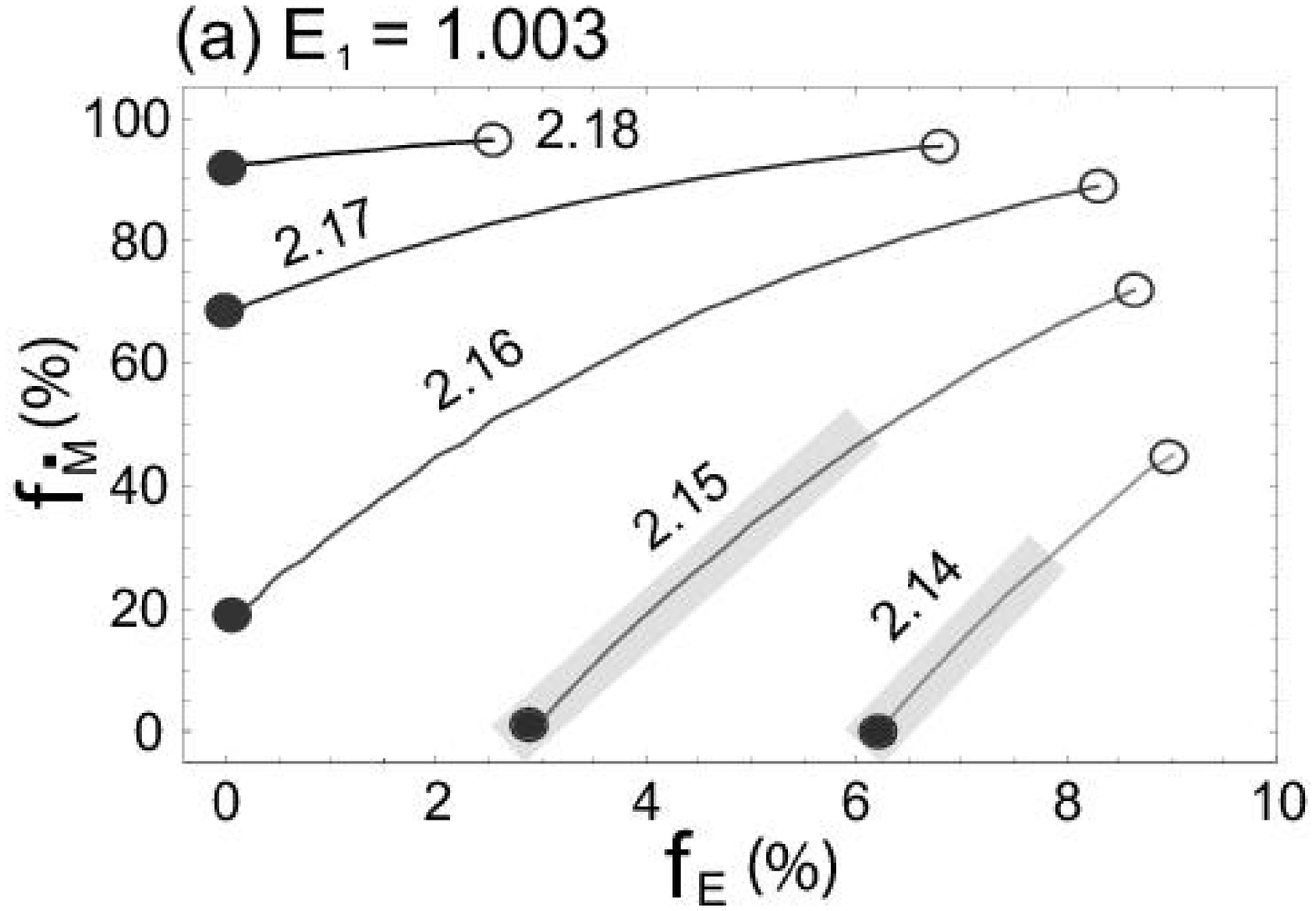}\plotone{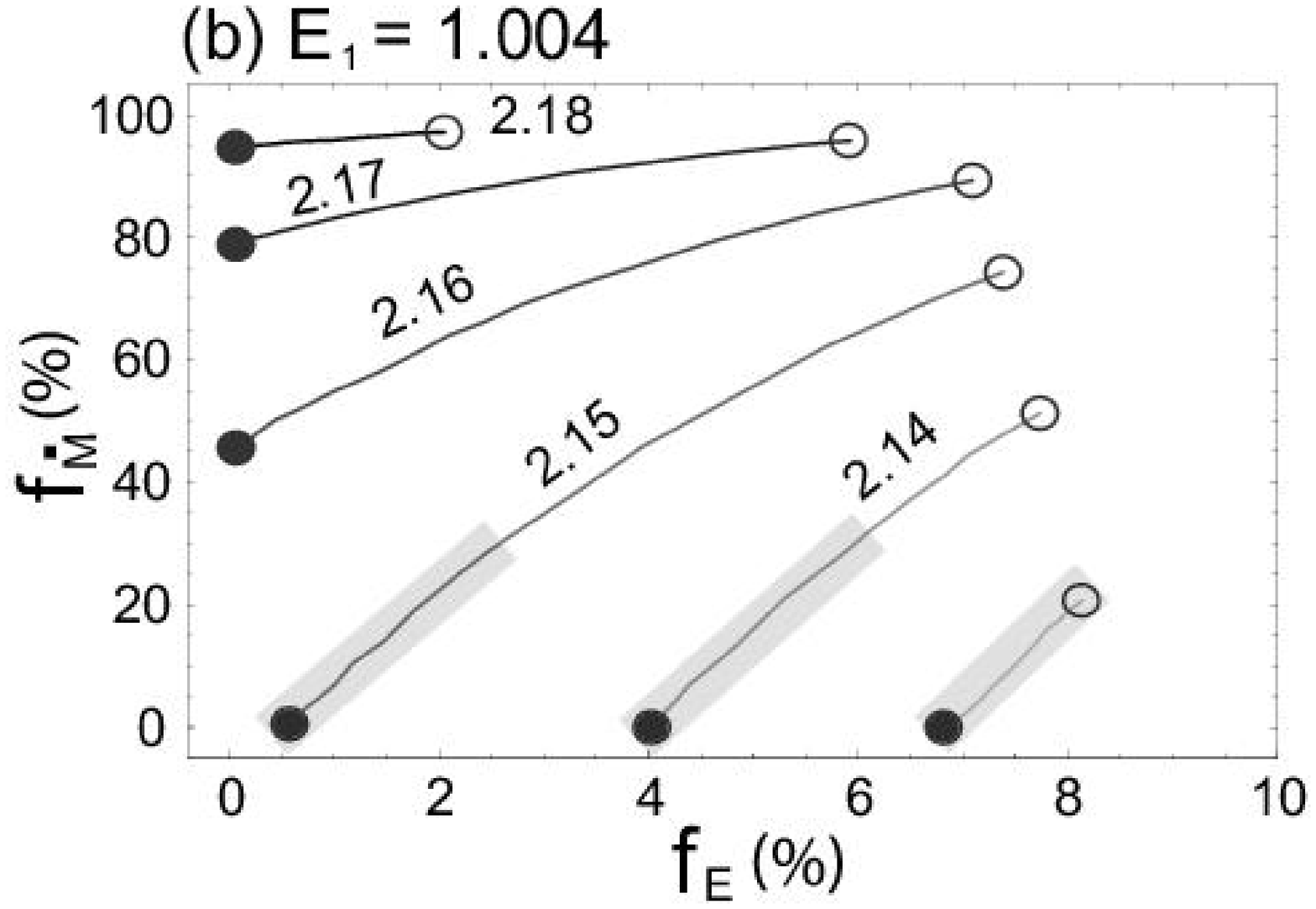}\plotone{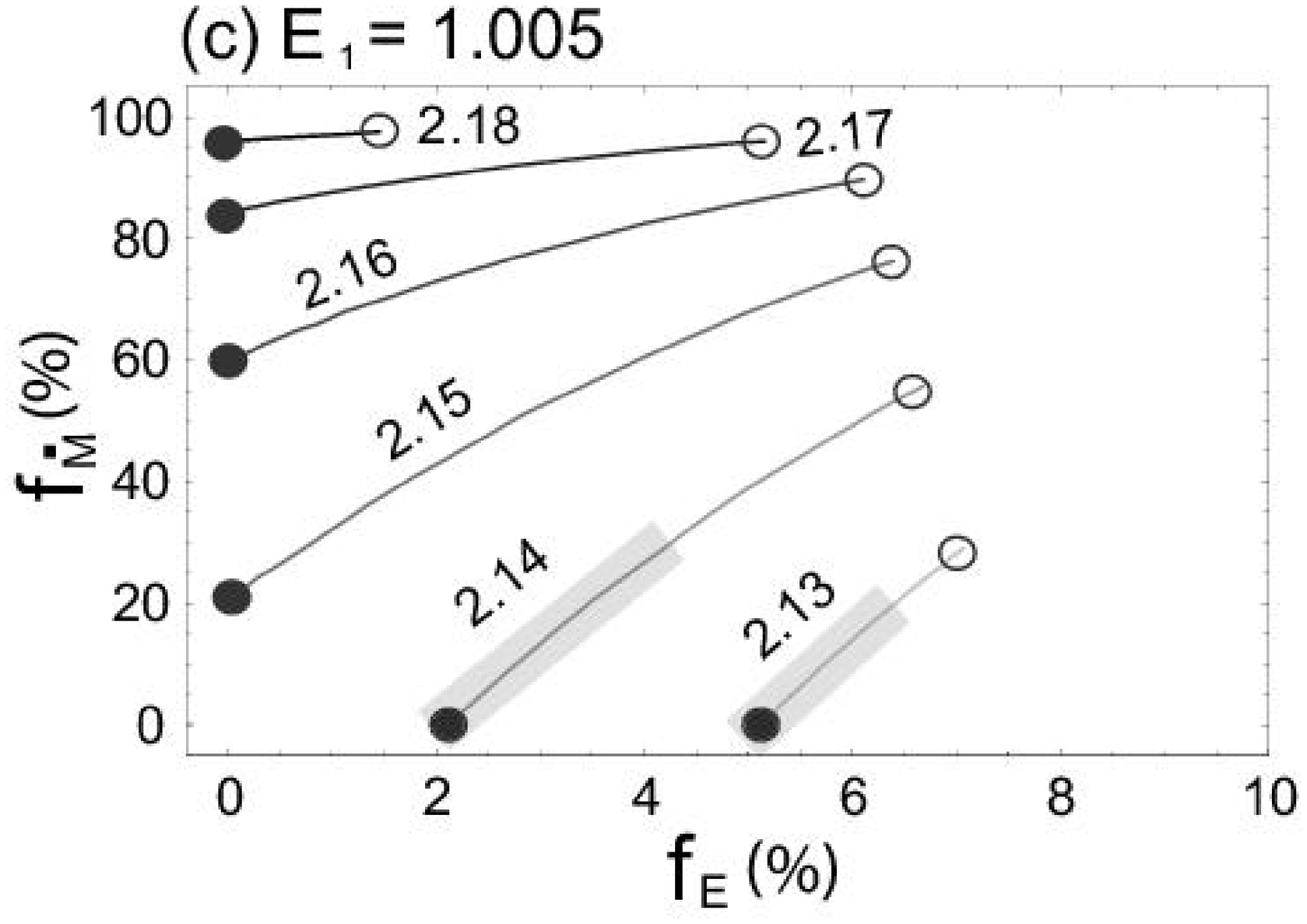}
\caption{Same as Fig.~\ref{fig:E-a0} but for $a/m=0.99$ case.}
\label{fig:E-a099}
\end{figure}

Figures~\ref{fig:E-a0} ($a/m=0$) and \ref{fig:E-a099} ($a/m=0.99$)
display the same mass loss fraction $f_{\dot{M}}$ as a function of
the energy loss fraction $f_E$, corresponding to the solutions in
Figure~\ref{fig:rsh-a0} and \ref{fig:rsh-a099}, respectively. At a
first glance, there is an explicit positive correlation between
$f_E$ and $f_{\dot{M}}$. As the angular momentum increases, maximum
value of $f_E$ is clearly reduced. In these figures the upper left
portion corresponds to the outflow solutions with larger
$f_{\dot{M}}$ and smaller $f_E$ (i.e., less energetic outflows)
while the lower right portion represents the solutions with smaller
$f_{\dot{M}}$ and larger $f_E$ (i.e., more energetic outflows).
Therefore, the results above suggest that more energetic outflowing
particles may be separated from a shock front when the upstream flow
possesses smaller angular momentum regardless of black hole spin
$a$. Mass loss fraction $f_{\dot{M}}$ can become as high as
$\lesssim 95\%$ for both $a/m=0$ and $a/m=0.99$ cases regardless of
energy $E_1$, whereas energy loss fraction is only $f_E < 1\%$ for
$a/m=0$ case but $f_E \lesssim 10\%$ for $a/m=0.99$ case. By direct
comparison of these figures, we note that rotation of a black hole
$a$ is also effective such that $d f_E / d f_{\dot{M}}(a/m=0.99) > d
f_E / d f_{\dot{M}}(a/m=0)$, i.e. an increase in mass loss would
allow more energy loss for $a/m=0.99$ case. Therefore, more
energetic outflows can be expected from a shock around a rotating
black hole.
%
%Lastly, we note that more input energy $E_1$ consistently yields
%smaller energy loss $f_E$ but larger mass loss $f_{\dot{M}}$; i.e.,
%less energetic outflows.

 % -------------------------------------

 % -------------------------------------

\begin{figure}[t]% ------------------------------------- Figure~6
\epsscale{1}\plottwo{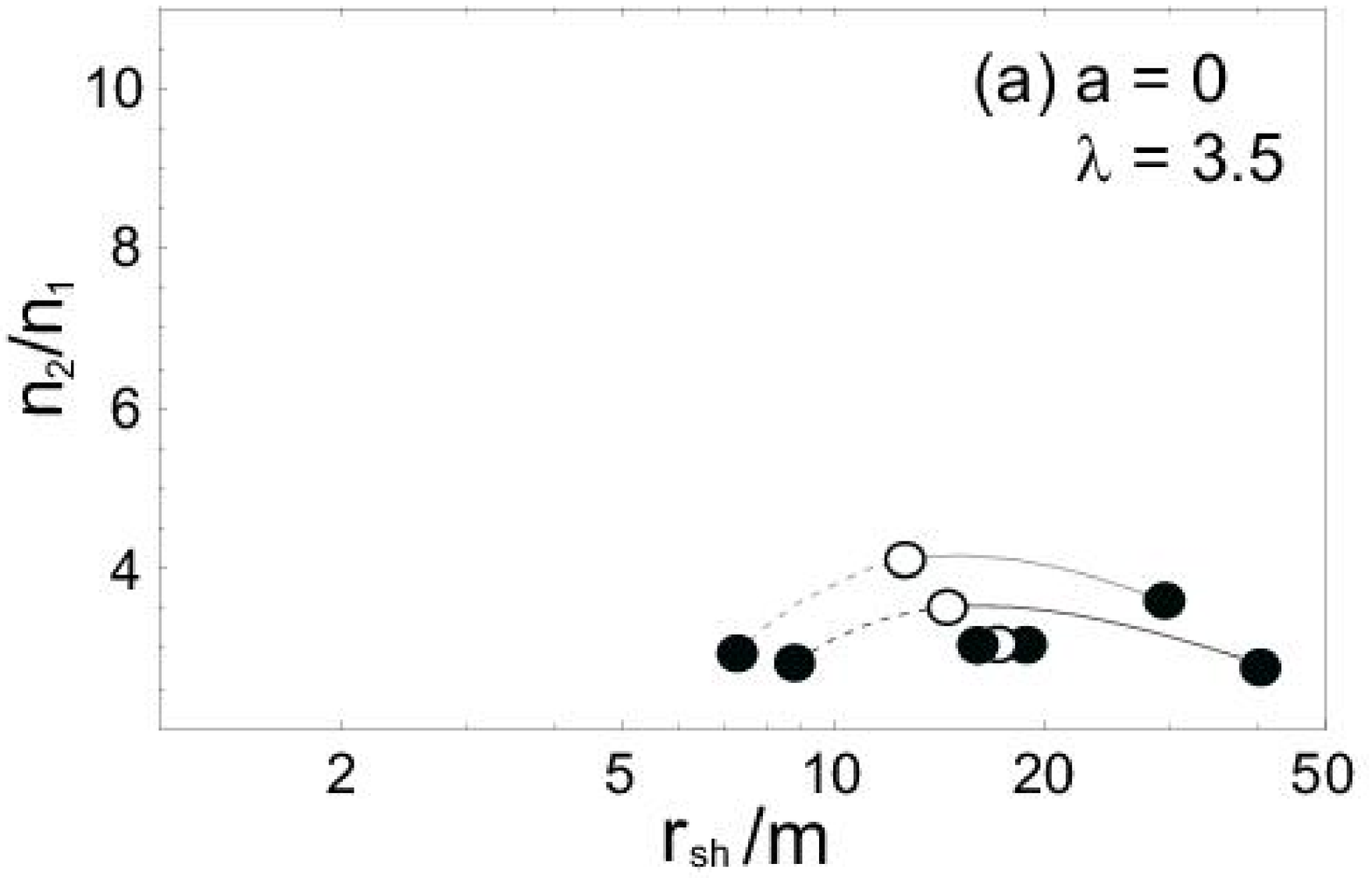}{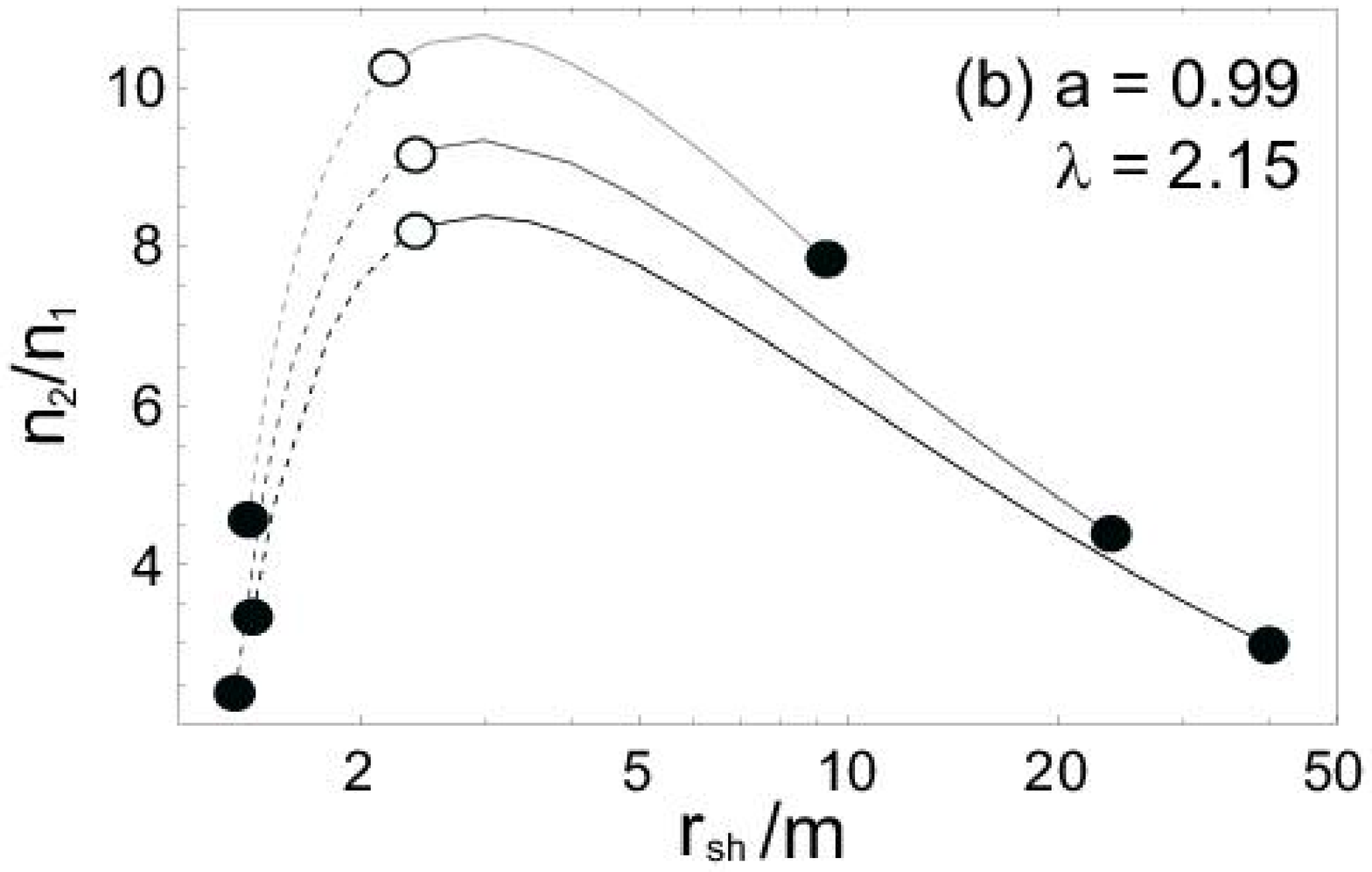} \caption{$n_2/n_1$ vs.
$r_{\rm sh}$ for (a) $a/m=0$ and (b) $0.99$ cases. We choose
$E_1=1.003,~1.004,~1.005$ from top to bottoms curves. Other
notations are the same as before. } \label{fig:n12}
\end{figure}

The upstream flow energy $E_1$ also affects the mass loss fraction.
It is noted that the energy-dependence of $f_{\dot{M}}$ is strong;
higher energy $E_1$ can lead to larger mass outflow fraction for a
given angular momentum $\lambda$ and a shock location $r_{\rm sh}$.
For instance, with $\lambda = 3.5$ for $a/m=0$ case (in
Fig.~\ref{fig:E-a0}), $f_{\dot{M}} < 20\%$ when $E_1=1.003$, $20-30
\%$ when $E_1=1.004$ and as large as $\sim 40\%$ when $E_1=1.005$.
With $\lambda=2.16$ for $a/m=0.99$ (in Fig.~\ref{fig:E-a099}),
$f_{\dot{M}} \sim 20-85 \%$ when $E_1=1.003$, $40-85\%$ when
$E_1=1.004$, and it is as high as $60-85\%$ when $E_1=1.005$. As a
reference, we also examine the energy-dependence of the shock
strength in Figure~\ref{fig:n12} where the (local) compression ratio
$n_2/n_1$ is plotted against $r_{\rm sh}$ for different energies:
$E_1=1.003,~1.004$ and $1.005$ as used before. Angular momentum
$\lambda$ is fixed in each case to see energy-dependence alone
(although different values of $\lambda$ for different spin $a$ must
be chosen to obtain the solutions). We find that our dissipative
shocks (coupled to mass loss) become stronger with decreasing
energy, a behavior similar to the typical types of shock formation
(e.g., see Lu et al. 1997 for adiabatic shocks; Lu \& Yuan 1998,
Fukumura \& Tsuruta 2004 for isothermal shocks).

 % -------------------------------------

\begin{figure}[t]% ------------------------------------- Figure~7
\epsscale{0.5} \plotone{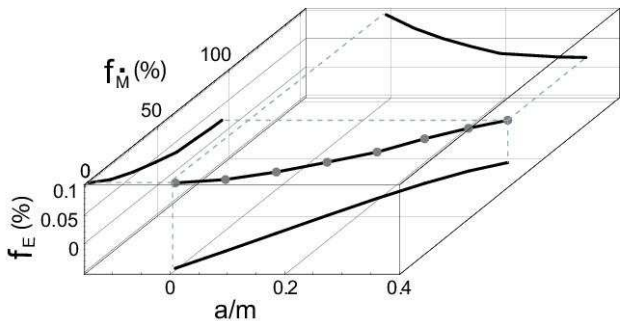} \caption{Mass loss efficiency
$f_{\dot{M}}$ as a function of black hole spin $a$ and energy loss
efficiency $f_E$ for a set of fixed parameters. We choose
$\lambda=3.45$ and $r_{\rm sh}/m=30$. Solution curve is projected on
each plane as shown. } \label{fig:spin}
\end{figure}

To exclusively illustrate the black hole spin dependence $a$ of the
mass loss efficiency $f_{\dot{M}}$ we fix all other parameters
($E_1, \lambda, r_{\rm sh}$) except for $a$. Figure~\ref{fig:spin}
shows $f_{\dot{M}}$ against $a$ and $f_E$ for $\lambda=3.45$ and
$r_{\rm sh}/m=30$. As seen in the earlier results, black hole
rotation alone can clearly enhance the efficiency of mass outflows
$f_{\dot{M}}$ from $\sim 3\%$ (for $a/m=0$) up to $\sim 95\%$ (for
$a/m=0.35$). On the other hand, the corresponding energy loss
efficiency $f_E$ remains as low as $\sim 0.02 - 0.1\%$. Note here
that $\lambda$ would have to be properly adjusted in order to obtain
the solutions for higher black hole spin $a$.

 % -------------------------------------

\begin{figure}[t]% ------------------------------------- Figure~8
\epsscale{0.5} \plotone{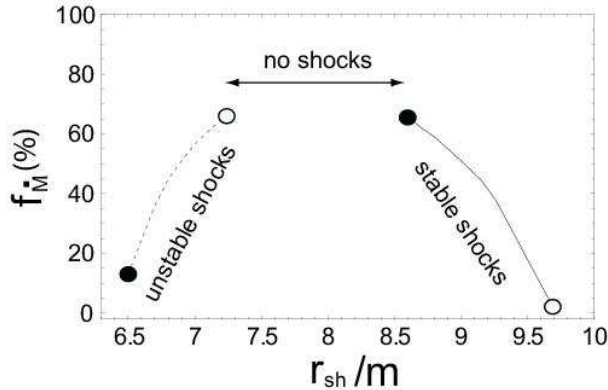} \caption{Mass loss efficiency
$f_{\dot{M}}$ vs. $r_{\rm sh}$ for $a/m=0$. We set $E_1=1.000001$
and $\lambda=3.73$. Notations are the same as in
Figure~\ref{fig:rsh-a0}. } \label{fig:low-E}
\end{figure}

We have chosen above some representative values for the flow energy
for parametric purpose. Weakly viscous/inviscid accretion in general
is a good model for some limited specific cases, like our Galactic
center, for example. For such specific cases, the realistic choice
of energy should be very small. From this perspective we examine to
see whether low energy flows can still produce shock-driven
outflows. Figure~\ref{fig:low-E} shows mass loss efficiency
$f_{\dot{M}}$ as a function of $r_{\rm sh}$ for $a/m=0$. We set
$E_1=1.000001$ and $\lambda=3.73$. Mass outflows can indeed be
produced with $f_{\dot{M}}$ ranging from $\sim 1\%$ up to $\sim
65\%$. Both unstable and stable shocks are present as in the earlier
cases but not continuously connected (no shock regions between the
two). The range of shock location is much narrower in radius in this
case, over which the mass loss efficiency can significantly change
as mentioned above. We will discuss this more in the Discussion
section.

 % -------------------------------------

The major correlations we find among the primary parameters are
summarized in Table~\ref{tab:tbl-1}. Table~\ref{tab:tbl-2} shows
various correlations with shock strength. Compression ratio
$n_2/n_1$ is strongly correlated with ($r_{\rm sh}, f_E,
f_{\dot{M}}$); stronger shocks are expected in regions very close to
the central engine particularly around a rotating black hole. Strong
shocks in principle are accompanied by high energy and mass loss
fractions.

\begin{deluxetable}{c|cccc}%[h]% ------------------------------- Table~1
\tabletypesize{\scriptsize} \tablecaption{Correlation among primary
quantities. \label{tab:tbl-1}} \tablewidth{0pt} \tablehead{Parameter
& $r_{\rm sh}$ & $f_E$ & $f_{\dot{M}}$ & $n_2/n_1$ } \startdata
          $\lambda$  & $+$ & $--$ & $++$  & $\leftrightarrow$   \\
          $E_1$ & $\leftrightarrow$ & $--$ & $+$  & $--$  \\
\enddata
\tablecomments{Strong positive (negative) correlation is denoted by
$++$ ($--$) while relatively weak positive (negative) correlation is
denoted by $+$ ($-$). No significant correlation is shown by
$\leftrightarrow$.}
\end{deluxetable}

\begin{deluxetable}{c|ccc}%[h]% ------------------------------- Table~2
\tabletypesize{\scriptsize} \tablecaption{Correlation with shock
strength $n_2/n_1$ among primary quantities. \label{tab:tbl-2}}
\tablewidth{0pt} \tablehead{Parameter & $r_{\rm sh}$ & $f_E$ &
$f_{\dot{M}}$  } \startdata
          $n_2/n_1$  & $--$ & $++$ & $++$   \\
\enddata
\tablecomments{Notations are the same as in Table~\ref{tab:tbl-1}.}
\end{deluxetable}

To sum up, our results show that strongest stable shocks generally
develop at the smallest radii (closer to the black hole),
accompanied by the largest mass loss $f_{\dot{M}}$ (and the largest
energy loss $f_E$ as well). The rotation of the black hole
apparently amplifies the shock strength by more than a factor of
two, also extending the outflowing site significantly inward (i.e.,
$r_{\rm sh}/m \gtrsim 2-3$ when $a/m=0.99$ while $r_{\rm sh}/m
\gtrsim 12$ when $a/m=0$). We will make some implications of the
obtained shock-outflow solutions in the last section, \S 4.

\subsection{Global Accreting Flows}

\begin{figure}[h]% ------------------------------------- Figure~9
\epsscale{1}\plotone{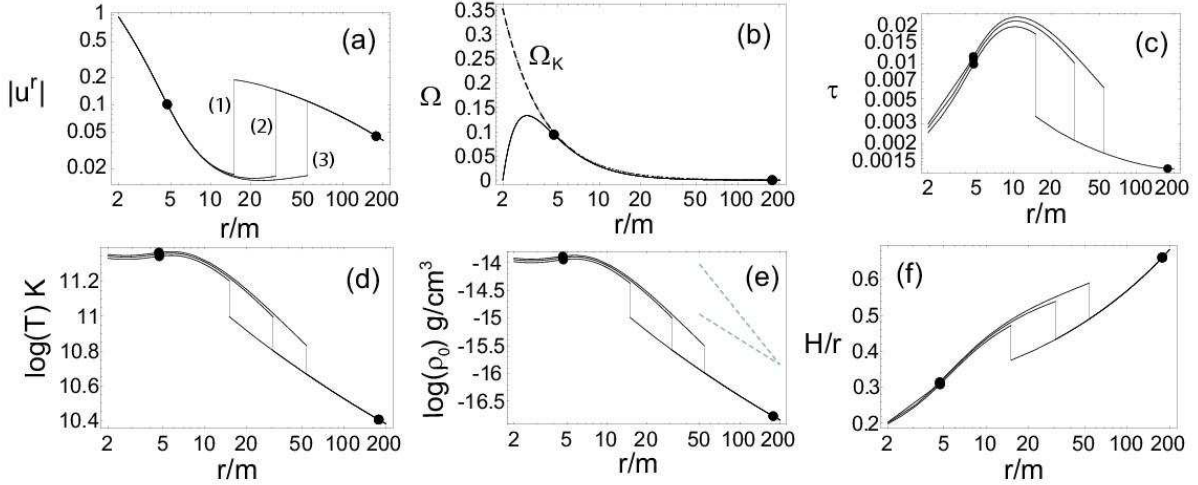} \caption{Global shock-included
solutions, (1)-(3), for $a/m=0$ case. See Table~\ref{tab:tbl-3} for
the parameters. (a) radial velocity $|u^r|$, (b) angular velocity
$\Omega$, (c) scattering optical depth $\tau$, (d) temperature $T$,
(e) density $\rho_0$, and (f) aspect ratio $H/r$. Vertical lines
denote shocks while filled circles indicate outer/inner critical
radii. In (b) Keplerian profile $\Omega_{K}$ is denoted by a dotted
curve. In (e) dotted gray lines show the slopes of $-3/2$ and $-3$.
} \label{fig:global-1}
\end{figure}

\begin{figure}[h]% ------------------------------------- Figure~10
\epsscale{1}\plotone{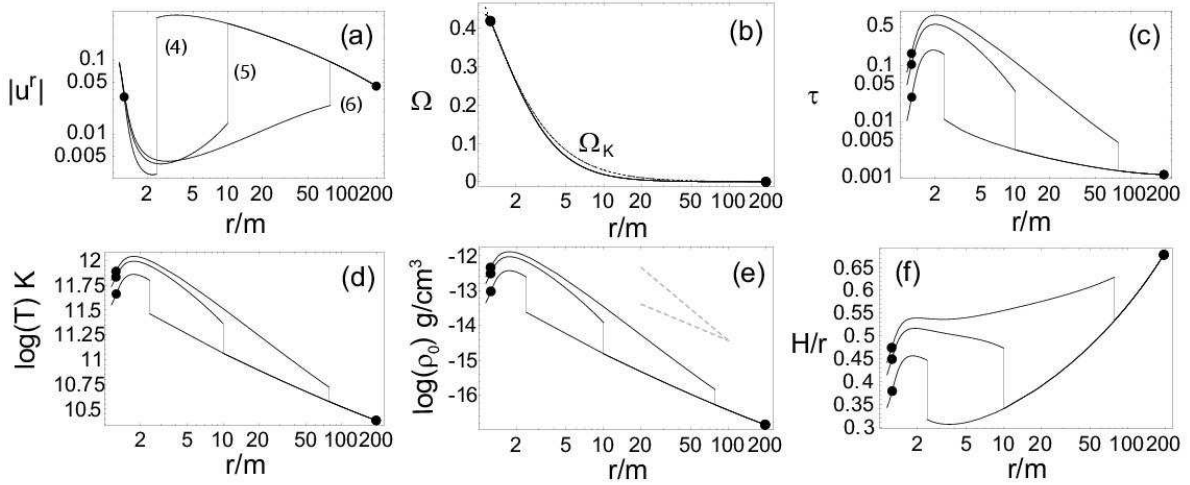} \caption{Same as
Figure~\ref{fig:global-1} but $a/m=0.99$ case. See
Table~\ref{tab:tbl-4} for the parameters.} \label{fig:global-2}
\end{figure}

Samples of the global run of the physical parameters of accretion
flows that include shocks are given in Figures~\ref{fig:global-1}
($a/m=0$) and \ref{fig:global-2} ($a/m=0.99$). In
Tables~\ref{tab:tbl-3} and \ref{tab:tbl-4} we provide the specifics
of the shocks associated respectively with the flows of
Figures~\ref{fig:global-1} and \ref{fig:global-2}. These plots have
been made assuming that all flows accrete at 1\% of the Eddington
rate, i.e. $\dot{m} \equiv \dot{M} / \dot{M}_{\rm Edd} = 0.01$ and
that the black hole mass is $m = 10^7 \Msun$, typical of Seyfert
nuclei. Each panel shows (a) the radial velocity $|u^r|$, (b) the
angular velocity $\Omega$, (c) the electron scattering optical depth
defined by $\tau \equiv n \sigma_T H$, (d) flow temperature $T$ [K],
(e) the density $\rho_0$ [g~cm$^{-3}$], and (f) the ratio of the
vertical scale-height to the radius $H/r$. Vertical lines denote the
positions of shocks connecting the upstream and downstream values of
the corresponding quantities of each flow. Dotted curve in panel (b)
shows the Keplerian angular velocity $\Omega_{K}$. Note that for
clarity purposes we only show three of the representative shock
solutions obtained, although the shock can occur at any radius
between the outermost and innermost solutions (with different values
of $f_{\dot{M}}$ and $f_E$).
%
%
%Sample solutions for shock-included, global accreting flows are
%displayed in Figures~\ref{fig:global-1} and \ref{fig:global-2}.
%
%\assume the radiative efficiency to be 1\% {\sf ???}[i.e.,
%$\eta_{\rm eff} \equiv L_{Edd}/(m_p c^2) = 0.01$ where $L_{Edd}$ is
%the Eddington luminosity]. We also take dimensionless mass-accretion
%rate to be $\dot{m} \equiv \dot{M}/\dot{M}_{\rm Edd} = 0.01$,
%relevant for typical Seyfert nuclei.
%
%
%Figure~\ref{fig:global-1} shows a global solution for the $a/m=0$
%case with $E_1=1.003, \ell=3.5$. We find $r_{sh}/m=21$ and
%$f_{\dot{M}}=0.064$. Each panel shows (a) the radial speed
%$|v^r/c|$, (b) the angular velocity $\Omega$, (c) the scattering
%optical depth $\tau \equiv n \sigma_T H$, (d) teh flow temperature
%$\log T$ [K], (e) the number density $\log n$ [cm$^{-3}$], and (f)
%the normalized vertical scale-height $H/r$. Vertical line denote a
%shock connecting upstream and downstream flows. Dotted curves
%represent the sound speed in panel (a) while Keplerian angular
%velocity $\Omega_K$ in panel (b).

%[Demos: modified below by KF as of 3/21]

We find that in both the non-rotating (Fig.~\ref{fig:global-1}) and
rotating (Fig.~\ref{fig:global-2}) cases the upstream flow density
scales as $\rho_{0,1}(r) \sim r^{-3/2}$, as in the ADAF self-similar
solution. On the other hand, as we explain below, the downstream
flow density generally has a slightly steeper power-law slope,
$\rho_{0,2}(r) \sim r^{-3/2} - r^{-3}$ unless the shock location is
too close to the horizon. As seen in Figures~\ref{fig:global-1}a and
\ref{fig:global-2}a, this is primarily because of the decreasing
(radial) downstream flow speed $|u^r_2(r)|$, whose feature is more
obvious around a rotating black hole. After the shock transition,
frame-dragging of the rotating black hole forces the accreting flow
to accelerate more in the toroidal direction (i.e., corotate) rather
than radial direction. Under adiabatic assumption, at the same time,
the flow temperature continues to rise (Fig.~\ref{fig:global-1}d and
\ref{fig:global-2}d) due to $p dV$ compression of the flow. Since
$c_s \propto T^{1/2}$, sound speed also increases. The resulting
(thermal) pressure gradient ($dP/dr<0$) prevents the incoming flow
from speeding up and in fact decelerate the downstream flow (for a
while) until the flow falls deep inside the gravitational potential
well to radially speed up again. Consequently, the fluid motion is
governed mainly by increasing toroidal velocity rather than radial
one (compare Fig.~\ref{fig:global-1}a with
Fig.~\ref{fig:global-2}a). This results in a larger radial velocity
gradient $d |u^r_2| / dr>0$ in Figure~\ref{fig:global-2}a.

In differentially-rotating flows, the Keplerian frequency increases
with decreasing $r$ even more rapidly than the sound speed, leading
to a decreasing scale-height $H$ [see equation~(\ref{eq:H})]. The
dependence of these two quantities, $\Omega_K$ and $c_s$, on the
radius in the downstream flow could lead to geometrically slim/thick
all the way to the horizon [see, e.g., $H/r \gtrsim 0.2 - 0.3$ in
Figs.~\ref{fig:global-1}f and \ref{fig:global-2}f].

Recalling $\rho_0 \propto 1 /(r H |u^r|)$ from
equation~(\ref{eq:mdot1}), the above fact that both $|u^r_2|$ and
$H$ decrease allows a very steep downstream density profile
$\rho_{0,2}(r)$ after the shock transition (and this is more so in a
rotating black hole case because of larger drop in $|u^r_2|$ as
mentioned earlier). The downstream flows become slightly more opaque
to electron scattering because of the shock compression, yet they
continue to remain optically-thin ($\tau < 1$).
%
%
%{\sf [ Keigo the explanation you give does not quite work. The
%dependence of the density on the radius has more do to with the
%radial velocity which you do not show in the figs. the speed of
%sound is proportional to the square root of the temperature which
%increases with $r$ as expected. However the radial velocity
%DECREASES for a while after the shock and that is the gives rise to
%the much steeper density profile. This seems to be the case mainly
%in the $a=0.99$ case. We need to rephrase this section. the velocity
%presumably decreases because of the increase in the temperature due
%to pdV compression of the flow. Also the geometric cross section of
%the flow maybe different than spherical near the horizon]}.
%
%
%Because the radial speed (equivalent to Mach number {\sf [not
%really; as i pointed out above, the sound speed also increases
%inward so that the decrease in the Mach number is partly due to the
%increase in the temperature. it is this that apparently chokes the
%flow and slows it down until eventually the infinite grav. potential
%of the BH wins over and sucks the fluid in}]) after the shock tends
%to decrease (for a while) whereas Keplerian angular velocity keeps
%rising, density profile can be significantly steep in the downstream
%region (sound speed only slightly increases).
%
%

\begin{deluxetable}{c|cccc}%[h]% ------------------------------- Table~3
\tabletypesize{\scriptsize} \tablecaption{Sets of Parameters for the
Global Solutions in Figure~\ref{fig:global-1}. \label{tab:tbl-3}}
\tablewidth{0pt} \tablehead{Model & $r_{\rm sh}/m$ & $n_2/n_1$ &
$f_E$ (\%) & $f_{\dot{M}}$ (\%) } \startdata
(1) & $15$ & $4.2$ & $0.319$ & $52.4$   \\
(2) & $31$ & $3.8$ & $0.136$ & $47.3$   \\
(3) & $54$ & $3.1$ & $0.00253$ & $43.4$   \\
\enddata
\tablecomments{Note that $a/m=0,~E_1=1.003,~\lambda=3.6$.}
\end{deluxetable}

\begin{deluxetable}{c|cccc}%[h]% ------------------------------- Table~4
\tabletypesize{\scriptsize} \tablecaption{Sets of Parameters for the
Global Solutions in Figure~\ref{fig:global-2}. \label{tab:tbl-4}}
\tablewidth{0pt} \tablehead{Model & $r_{\rm sh}/m$ & $n_2/n_1$ &
$f_E$ (\%) & $f_{\dot{M}}$ (\%) } \startdata
(4) & $2.4$ & $10.2$ & $8.3$ & $89$   \\
(5) & $10$ & $7.9$ & $2.6$ & $51.0$   \\
(6) & $78$ & $2.73$ & $0.002$ & $17.6$   \\
\enddata
\tablecomments{Note that $a/m=0.99,~E_1=1.003,~\lambda=2.16$.}
\end{deluxetable}

 % -------------------------------------

 % -------------------------------------

\section{Discussion \& Conclusions}

In this work we have examined the structure of accretion flows that
includes shocks of more general character than those discussed so
far in the literature \citep[e.g.,][]{Cha90}. In particular we have
examined whether it is possible to have flows with shocks at which
part of the mass and/or energy fluxes are lost and do not
participate in the shock transition (this maybe the case in
multidimensional shocks or in shocks that involve acceleration of
particles that escape from the shock region). Such generalized
shocks obey jump conditions more general than those of
Rankine-Hugoniot and it is not {\it a priori} certain if accretion
flows can allow for such shocks. We have also examined the degree of
mass and energy loss allowed that are at the same time consistent
with the continuation of the downstream flow through another sonic
transition onto the black hole. In this respect we have focused our
study on shocks in which the energy per particle for those escaping
the shock transition is greater than that of the local gravitational
potential. In such a case one can argue that these particles can
escape to infinity producing the jets/winds observed in many
accretion-powered systems. We have found that there are indeed flows
with global parameters (i.e. $\lambda$ and $E$) that allow for such
shocks. On the other hand, we have also obtained solutions at which
the energy per particle of the escaping matter is smaller than the
local gravitational potential, indicating that in all likelihood
these particles will stagnate and eventually join the rest of the
flow onto the compact object.
%
%{\sl \{KF: We examined the mass loss efficiency that are coupled to
%the shock solutions for various flows. Most importantly, in order
%for the outflowing particles to leave the shocked flow surface, it
%must posses sufficiently large kinetic energy exceeding some
%threshold which is determined by equating outward force against
%inward gravitational pull. Otherwise, the released particles might
%stagnate around the shocked flow surface or is advected with the
%downstream flow, not being ejected away\}}.
%
However, a deeper understanding of the precise mechanisms that can
lead to the particles that do not participate in the shock
transition, as we conjectured above, requires the knowledge of the
detailed microphysics of the shock (i.e. the acceleration
efficiency, the fraction of energy put into relativistic particles
at the shock and their transport in and around the shock geometry);
these are beyond the scope of the present paper.

To get a crude estimate of the energetics of these outflowing
particles, we compute a threshold energy of the escaping particles
under a set of assumed parameters. The total energy flux carried by
the escaping particles (of energy $\Delta E$) is given by $\Delta E
\times 4\pi r_{\rm sh} \bar{H} \bar{n} \bar{u}^r$ [erg~s$^{-1}$]
where $\bar{H} \equiv (H_1+H_2)/2, \bar{n} \equiv (n_1+n_2)/2$ and
$\bar{u}^r \equiv (u^r_1+u^r_2)/2$ are evaluated at the shock
location ($r=r_{\rm sh}$). The minimum kinetic energy necessary for
the separated particles (of mass outflow rate $\Delta \dot{M}$) to
escape the bulk flow is given by $\Delta \dot{M} v_{c}^2 /2$
[erg~s$^{-1}$] where $v_{c}$, to be determined below, is the
required critical velocity of the particles. By equating these, we
solve for $v_{c}$ to obtain
\begin{eqnarray}
v^2_{c} \sim 8 \pi r_{\rm sh} \bar{H} \bar{n} \bar{u}^r m_p c^2
\Delta E / \Delta \dot{M} \ .
\end{eqnarray}
For the outflowing particles to be effectively decoupled from the
bulk flow, kinetic energy associated with this critical velocity
should exceed or be at least comparable to the corresponding
gravitational binding energy at the shock location. This condition
reads
\begin{eqnarray}
\frac{1}{2} v_c^2 \gtrsim \frac{G m}{r_{\rm sh}} \ ,
\label{eq:condition-1}
\end{eqnarray}
or
\begin{eqnarray}
\Delta E \gtrsim E_c \ , \label{eq:condition-2}
\end{eqnarray}
where $E_c \equiv G m \Delta \dot{M} / (4 \pi r_{\rm sh}^2 \bar{H}
\bar{n} \bar{u}^r m_p c^2)$.
%
%
%The energy flux carried with the escaping particles is given by
%$\Delta E \times 4\pi r_{\rm sh} \bar{H} \bar{n} \bar{u}^r$
%erg~s$^{-1}$ where $\bar{H} \equiv (H_1+H_2)/2, \bar{n} \equiv
%(n_1+n_2)/2$ and $\bar{u}^r \equiv (u^r_1+u^r_2)/2$ are evaluated at
%the shock location $r_{\rm sh}$. The corresponding mass-energy flux
%of the escaping particles is $\Delta \dot{M} v^2_{esc}$ erg~s$^{-1}$
%where $v_{esc}$ is the escape velocity of the particles. By equating
%these, we obtain
%%
%\begin{eqnarray}
%v^2_{esc} \sim 4\pi r_{\rm sh} \bar{H} \bar{n} \bar{u}^r m_p c^2
%\Delta E / \Delta \dot{M} \ .
%\end{eqnarray}
%
%{\sf Then the condition for these particles to escape reads}
%
%\begin{eqnarray}
%\Delta E \gtrsim E_c \ ,
%\end{eqnarray}
%%
%where $E_c \equiv G m \Delta \dot{M} / (2\pi r_{\rm sh}^2 \bar{H}
%\bar{n} \bar{u}^r m_p c^2)$.
%
%
Since all these parameters ($\bar{H}, \bar{n}, \bar{u}^r$) are
functions of shock location $r_{\rm sh}$ which depends on
$f_{\dot{M}}$, $E_c$ is obtained once we specify $f_{\dot{M}}$. Now,
we can review the shock solutions obtained earlier from this new
perspective: In Figures~\ref{fig:rsh-a0}-\ref{fig:E-a099} the
allowed shock solutions satisfying $\Delta E \ge E_c$, which is
relevant for outflows that may escape the shocked flow, are
displayed by shaded regions, while the rest of the solutions
correspond to the outflows presumably bound to the bulk flow.
According to these estimates, an outflow that expels a certain
amount of the preshock accreting matter is possible, over a rather
broad range of radii in the vicinity of the central object;
%
%{\sl \{ In this estimate certain degrees of outflows could be
%possible over a wide range of radius not too far from the central
%engine\} }
%
up to nearly 10\% of matter may participate in producing seed
particles for jets/winds within $\sim 20 - 40$ gravitational radii
around a non-rotating black hole, while as much as 50\% of preshock
matter could contribute to forming jets/winds within $\sim 2 - 30$
gravitational radii around a rapidly-rotating black hole. More
interestingly, our solutions also allow for dissipative shocks ($f_E
\ne 0$) with negligible mass loss ($f_{\dot{M}} \sim 0$). Such
solutions that involve the loss of infinitesimal mass and finite
amount of energy can lead to relativistic outflows from the
corresponding shocks.
%
%{\sl \{ lost and These solutions correspond to a case where all the
%liberated energy is transferred to a small population of nonthermal
%particles \} }.
%
Perhaps, outflows via this type of shocks may produce kinematically
strong jets (with high Lorentz factor) rather than moderate velocity
winds. Thus, shock-driven outflows may provide clues on the origin
of the jets/winds from the inner accretion regions, for instance, in
some active galaxies, e.g., M87 and 3C120.

%\hspace{10in}

In this simple model, shocks in principle could drive axisymmetric
outflowing matter.
Near the base of the outflow, as illustrated in
Figure~\ref{fig:schematic}, its shape would be that of a hollow
cone. However, as it propagates to longer distances we do not expect
that this shape would be preserved all the way.
%
%Because we treat the thickness of a shock as infinitesimally small
%(i.e. mathematical discontinuity), it is difficult to discuss the
%actual (particle) acceleration mechanisms at a shock front,
%necessary for constraining the geometry of the outflow.
%
Within the framework of our current scenario it can be speculated
that the most of the outflows discussed here could be in a diffuse
form, while some strong outflows with low $f_{\dot{M}}$ and high
$f_E$ might possess a relatively more collimated geometry (perhaps
in the poloidal direction) due to its high Lorentz factor. In the
presence of large-scale magnetic fields the dissipated plasma would
stream along the field lines, producing collimated outflows.
However, it is beyond the scope of our model to further discuss the
exact geometry of the outflows.

%When efficient shock acceleration is in operation at a shock front,
%the estimated velocity $\bar{u}^r$ above could be much larger. In
%that case, the kinetic energy of the outflows would easily exceed
%the threshold value in the present estimate. Thus our estimate here
%may put a tighter constraint on the allowed shock-related quantities
%presented in Figures~\ref{fig:rsh-a0}-\ref{fig:E-a099}.

It should be noted that for a given $\lambda$ there exists the
degeneracy of shock-outflow solutions: i.e., for a single value of
angular momentum $\lambda$ there is a finite range of possible shock
locations allowed ($r^{\rm min}_{\rm sh} < r_{\rm sh} < r_{\rm
sh}^{\rm max}$) that corresponds respectively to outflows with
$f_{\dot{M}}^{\rm min} \le f_{\dot{M}} \le f_{\dot{M}}^{\rm max}$
where the indices ``min" and ``max" denote minimum and maximum
values, respectively. In the context of our model, one can predict
the shock location $r_{\rm sh}$ by the (observational) knowledge of
$(\lambda, E_1, f_{\dot{M}})$, although the obtained parameter space
can be altered by additional physical ingredients (e.g., viscosity,
magnetic fields, for instance).

Another issue to be addressed is the fact that in general critical
points are not necessarily equivalent to sonic points depending on
the flow geometry and equation of state used
\citep[e.g.,][]{Das07b}. That is, there could be a potential danger
that a false shock could occur in subsonic region between a critical
radius and an actual sonic radius, in which case the obtained shock
would be unphysical. To ensure that a physically valid shock forms
in supersonic regions, we have checked the validity of our shock
solutions by computing a three-velocity component of the flow
measured by a suitable local observer at the shock location. We
calculated the (radial) flow velocity $v^r$ measured by a locally
stationary observer in the corotating reference frame
\citep[e.g.,][]{Lu86, Lu98}
\begin{eqnarray}
v^r \equiv \frac{u_r u^r}{1+u_r u^r} \ , \label{eq:vr}
\end{eqnarray}
and compared this velocity to the local sound velocity $c_s$ given
by equation~(\ref{eq:sound}). All the shock solutions presented here
are found to form in supersonic regions (i.e., $|v^r| > c_s$ at
$r=r_{\rm sh}$).

We showed that low energy flows can still produce mass outflows with
suitable angular momenta. Although continuous accreting solutions
(i.e. shock-free solutions) are persistently present even for
smaller energy $E_1$ (as expected), we do not find shocked flow
solutions (or perhaps they are present but in much narrower
parameter space). It can be speculated that maybe by prohibiting our
jump condition in energy (energy dissipation at a shock front) we
may obtain shock-driven outflows for smaller flow energy $E_1$,
which could be a case similar to \citet{Das99} where they considered
little energy loss and found mass outflows in pseudo-Newtonian
geometry. However, powerful outflows should carry away a significant
amount of (kinetic) energy. Hence, outflow solutions should in
principle be coupled to energy dissipation as treated here.

We have explored a coupling between shock solutions in accretion and
mass/energy losses (fractions) under a scenario that the
shock-driven outflowing particles may participate in forming a base
of jets/winds. For various flow parameters with a given black hole
spin, we have shown, by steady-state, axisymmetric hydrodynamic
calculations, that the dissipative shock front could be a plausible
site where a fraction of the accreting matter can be decoupled as
jets/winds from the bulk accretion flows.

\acknowledgments

K.F. thanks John Cannizzo for his useful comments. We are also
grateful to the anonymous referee for several useful suggestions and
comments that clarified the manuscript. This research was supported
in part by an appointment to the NASA Postdoctoral Program at the
Goddard Space Flight Center, administered by Oak Ridge Associated
Universities through a contract with NASA.

%\clearpage


\begin{thebibliography}{999}
%\bibitem[Becker et al.(2001)]{Becker01} Becker, P. A., \& Subramanian, P., \& Kazanas,
%D. 2001, \apj, 552, 209
\bibitem[Begelman et al.(1984)]{BBR84} Begelman, M. C., Blandford, R. D., \& Rees, M. J. 1984, Rev. Mod.
Phys., 56, 255
\bibitem[Blandford \& Begelman(1999)]{BB99} Blandford, R. D., \& Begelman, M. C. 1999, \mnras, 303,
L1
%\bibitem[Blandford \& Ostriker(1978)]{Blandford78} Blandford, R. D., \& Ostriker, J. P. 1978,
%\apj, 221, L29
\bibitem[Blandford \& Payne(1982)]{BP82} Blandford, R. D., \& Payne, D. G. 1982, \mnras, 199,
883
\bibitem[Chakrabarti(1990)]{Cha90} Chakrabarti, S. K. 1990, Theory of Transonic Astrophysical Flows (World Scientific,
Singapore)
\bibitem[Chakrabarti(1996)]{Cha96} Chakrabarti, S. K. 1996, \mnras, 283, 325
\bibitem[Chakrabarti(1999)]{Cha99} Chakrabarti, S. K. 1999, \mnras, 351, 185
\bibitem[Chakrabarti \& Das(2004)]{Cha04} Chakrabarti, S. K., \& Das, S. 2004, \mnras, 349, 649

\bibitem[Contopoulos \& Lovelace(1994)]{Contopoulos94} Contopoulos, J., \& Lovelace, R. V. E. 1994, \apj, 429,
139
\bibitem[Contopoulos \& Kazanas(1995)]{Contopoulos95} Contopoulos, J., \& Kazanas, D. 1995, \apj, 441, 521
\bibitem[Das \& Chakrabarti(1999)]{Das99} Das, T. K., \& Chakrabarti, S. K. 1999, Classical
Quantum Gravity, 16, 3879
\bibitem[Das(2000)]{Das00} Das, T. K. 2000, \mnras, 318, 294
\bibitem[Das \& Chakrabarti(2007)]{Das07} Das, T. K., \& Chakrabarti, S. K. 2007, \mnras, 374,
729
\bibitem[Das(2007)]{Das07b} Das, T. K. 2007 (astro-ph/0704.3618)
\bibitem[Fender et al.(2004)]{Fender04} Fender, R. P., Belloni, T. M., \& Gallo, E.
2004 , \mnras, 355, 1105
\bibitem[Fukumura \& Tsuruta(2004)]{FT04} Fukumura, K., \& Tsuruta, S. 2004, \apj, 611, 964
\bibitem[Fukumura et al.(2007)]{Fukumura07} Fukumura, K., Takahashi, M., \& Tsuruta, S. 2007,
\apj, 657, 415
\bibitem[Junor et al.(1999)]{Junor99} Junor, W., Biretta, J. A., \& Livio, M. 1999, Nature, 401, 891
\bibitem[Kataoka et al.(2007)]{Kataoka07} Kataoka et al., 2007, (astro-ph/0612754)
\bibitem[Protheroe \& Kazanas(1983)]{Kazanas83} Protheroe, R. J., \& Kazanas, D. 1983, \apj, 265,
620
\bibitem[Kazanas \& Ellison(1986)]{Kazanas86} Kazanas, D., \& Ellison, D. C. 1986, \apj, 304, 178
\bibitem[Koide et al.(1999)]{Koide99} Koide, S., Shibata, K., \& Kudoh, T. 1999, \apj, 522,
727
\bibitem[K\"{o}nigl \& Kartje(1994)]{Konigl94} K\"{o}nigl, A., \& Kartje, J. F. 1994, \apj, 434, 446
\bibitem[Le \& Becker(2004)]{Becker04} Le, T., \& Becker, P. A. 2004, \apj, 617,
L25
\bibitem[Le \& Becker(2005)]{Becker05} Le, T., \& Becker, P. A. 2005, \apj, 632, 476
\bibitem[Livio(1999)]{Livio99} Livio, M. 1999, Phys. Rep., 311, 225
\bibitem[Lu et al.(1997)]{Lu97} Lu, J.-F., Yu, K. N., Yuan, F., \&
Young, E. C. M. 1997, \aap, 321, 665
\bibitem[Lu(1986)]{Lu86} Lu, J.-F. 1986, General Relativity and Gravitation,
18,?

\bibitem[Lu \& Yuan(1998)]{Lu98} Lu, J.-F., \& Yuan, F. 1998, \mnras, 295, 66
\bibitem[Lu et al.(1999)]{Lu99} Lu, J.-F., Gu, W.-M., \& Yuan, F. 1999, \apj, 523, 340
\bibitem[Mirabel et al.(1999)]{Mirabel99} Mirabel, I. F., \& Rodriguez, L. F. 1999, ARA\&A, 37, 409
\bibitem[Manmoto et al.(1997)]{Manmoto97} Manmoto, T., Mineshige, S., \& Kusunose, M.
1997, \apj, 476, 49
\bibitem[Narayan \& Yi(1994)]{Narayan94} Narayan, R., \& Yi, I. 1994, \apj, 428, L13
\bibitem[Nishikawa et al.(2005)]{Nishikawa05} Nishikawa, K.-I., Richardson, G., Koide, S.,
S hibata, K., Kudoh, T., Hardee, P., \& Fishman, G. J. 2005, \apj,
625, 60

\bibitem[Pelletier \& Pudritz(1992)]{Pelletier92} Pelletier, G., \& Pudritz, R. E. 1992, \apj, 394, 117

\bibitem[Subramanian et al.(1999)]{Subramanian99} Subramanian, P., Becker, P. A., \& Kazanas, D. 1999,
\apj, 523, 203
\bibitem[Sponholz \& Molteni(1994)]{Sponholz94} Sponholz, H., \& Molteni, D.
1994, \mnras, 271, 233
\bibitem[Takahashi et al.(2002)]{Takahashi02} Takahashi, M., Rilett, D., Fukumura, K., \& Tsuruta, S. 2002, \apj, 572,
950
\bibitem[Vlahakis et al.(2000)]{Vlahakis00} Vlahakis, N., Tsinganos, K., Sauty, C., \& Trussoni, E.
2000, \mnras, 318, 417
\bibitem[Yang \& Kafatos(1995)]{Yang95} Yang, R., \& Kafatos, M. 1995, \aap, 295, 238
\end{thebibliography}
\end{document}